\documentclass[%
reprint,
nofootinbib,
amsmath,amssymb,
aps,
]{revtex4-2}

\usepackage{graphicx}
\usepackage{dcolumn}
\usepackage{bm}

\usepackage{graphics}
\usepackage{amsmath,amssymb,braket}
\usepackage{mathtools}
\usepackage{fancyvrb,comment,array}
\usepackage{booktabs,multirow}

\allowdisplaybreaks[1]
\usepackage{color}
\newcommand{\tr}{\mathrm{tr}}

\newcommand{\La}{\mathcal{L}}
\newcommand{\OC}{\mathcal{O}}

\newcommand{\M}{\mathcal{M}}
\newcommand{\barDs}{\bar{D}^{(\ast)}}
\newcommand{\Xis}{\Xi^{(\ast)}_{cc}}
\newcommand{\Op}{\mathcal{O}}
\newcommand{\B}{\mathcal{B}}
\newcommand{\ex}{\mathrm{ex}}

\newcommand{\Slash}[1]{\ooalign{\hfil/\hfil\crcr$#1$}}

\def\simge{\mathrel{
\rlap{\raise~0.511ex \hbox{$>$}}{\lower~0.511ex \hbox{$\sim$}}}}
\def\simle{\mathrel{
\rlap{\raise~0.511ex \hbox{$<$}}{\lower~0.511ex \hbox{$\sim$}}}}
\def\bigs{\mathrel{
\rlap{\raise~0.531ex \hbox{$>$}}{\lower~0.531ex \hbox{$<$}}}}

\usepackage{booktabs}
\usepackage[normalem]{ulem}
\usepackage{multirow}

\usepackage{hyperref}

\begin{document}
\title{Possible $\barDs \Xis$ and $\Xis\Xis$ molecules as superflavor partners of $T_{cc}$}

\author{Manato~Sakai}
\email[]{msakai@hken.phys.nagoya-u.ac.jp}
\affiliation{Department of Physics, Nagoya University, Nagoya 464-8602, Japan}
\affiliation{Department of Physics, Tokyo Metropolitan University, Hachioji 192-0397, Japan}
\author{Yasuhiro~Yamaguchi}
\email[]{yamaguchi@hken.phys.nagoya-u.ac.jp}
\affiliation{Department of Physics, Nagoya University, Nagoya 464-8602, Japan}
\affiliation{Kobayashi-Maskawa Institute for the Origin of Particles and the Universe, Nagoya University, Nagoya, 464-8602, Japan}
\affiliation{Department of Physics, Tokyo Metropolitan University, Hachioji 192-0397, Japan}

\date{\today}

\begin{abstract}

    The doubly charmed tetraquark $T_{cc}$ has been reported by the LHCb experiment in 2022, and a lot of theoretical studies have been conducted. The small binding energy measured from the $D^{\ast + }D^0$ threshold indicates that $T_{cc}$ is a $DD^\ast$ molecule. 
    On the other hand, the superflavor symmetry, which relates heavy antiquarks to heavy diquarks, provides a useful framework for predicting the existence of partner exotic hadrons associated with $T_{cc}$.
    By replacing $\bar{D}^{(*)}$ with $\Xi_{cc}^{(*)}$ within this symmetry, $\barDs\Xis$ and $\Xis\Xis$ are expected to form partner structures of $T_{cc}$. 
    In this paper, we investigate bound and resonant states of $\barDs\Xis$ and $\Xis\Xis$ based on the one boson exchange potential, where $\pi$, $\rho$, $\omega$ and $\sigma$ are considered as bosons. The cutoff parameter and the coupling constants for $\barDs\Xis$ and $\Xis\Xis$ are taken to be the same as those for $T_{cc}$ due to superflavor symmetry.
    We also discuss the $\sigma$ coupling constant, which is uncertain, dependence of these mass spectra. A lot of bound and resonant states with some quantum numbers are obtained for each $\sigma$ coupling constant, but these mass spectra depend on the $\sigma$ coupling constant significantly.

\end{abstract}

\maketitle
\section{Introduction}\label{sec;introduction}

Recently, in the heavy quark sector, numerous exotic hadrons have been reported by some experiments such as Belle, LHCb, BABAR, BESIII, and so on~\cite{Belle:2003nnu,BaBar:2005hhc,Belle:2007dxy,Belle:2007hrb,BaBar:2008bxw,LHCb:2015yax,LHCb:2019kea,LHCb:2020bwg,LHCb:2020jpq,LHCb:2021vvq,LHCb:2021auc,LHCb:2021chn,BESIII:2020qkh}. 
An exotic hadron is a hadron that cannot be accommodated within the conventional $qqq$ and $q\bar{q}$ quark configurations.
Most of the exotic hadrons are considered to be multiquark systems, but these spacial configurations are not obvious.  
On the other hand, most exotic hadrons are found slightly below a threshold, thus these states are analyzed as loosely bound states by the molecular model~\cite{Tornqvist:1993ng,Guo:2017jvc}.

Hadron-hadron interactions are one of the remaining problems in the hadron physics. 
The interaction between hadrons is described using the effective model such as the one boson exchange potential~\cite{Machleidt:1987hj,Machleidt:2000ge} and quark-cluster model~\cite{Oka:1981ri,Oka:1981rj,Oka:2000wj,Sekihara:2023ihc}, or Lattice QCD calculation~\cite{Aoki:2009ji,Inoue:2010es,Ishii:2012ssm,Lyu:2023xro}.  
One of the “well-known'' hadron-hadron interactions is the one pion exchange potential (OPEP) between nucleons. 
The OPEP is a long-range interaction and plays a crucial role in binding a deuteron which is a loosely bound state of a neutron and proton with $I(J^P) = 0(1^+)$. 
In the deuteron system, the tensor force of the OPEP which induces the $S - D$ mixing provides a crucial attraction between nucleons.  
On the other hand, other boson exchange potentials play a important role in middle or short ranges. For instance, the one $\rho$ and $\omega$ exchange potentials are middle-range forces. 
In addition, the one $\sigma$ exchange potential also provides the middle range forces, but this mechanism has not been understood completely.  In fact, this coupling constant is not determined. 

Symmetries related to heavy quarks play an important role in exotic hadrons composed of heavy quarks. 
In particular, two symmetries are relevant: the heavy quark symmetry (HQS)~\cite{Neubert:1993mb,Grinstein:1995uv,Casalbuoni:1996pg} and the superflavor symmetry, also referred to as the heavy quark–antidiquark symmetry~\cite{Georgi:1990ak,Savage:1990di,Fleming:2005pd}.
In the heavy quark limit, the spin-spin interaction of heavy quarks is suppressed at order $\OC(1/m_Q)$. 
As a consequence, the heavy quark spin is conserved, leading to the emergence of the HQS. 
At the hadron level, the HQS leads to an approximate mass degeneracy between heavy hadrons with different spins. 
For example, the mass difference between the charmed pseudoscalar meson $D$ and charmed vector meson $D^\ast$ is about $140\,\mathrm{MeV}$, which reflects the approximate mass degeneracy between these charmed hadrons. 
This mass degeneracy induces coupled channel effects such as $DD-DD^\ast-D^\ast D^\ast$ mixing. This mixing provides the one pion exchange potential which is prohibited in the $DD\pi$ vertex because of the parity conservation. 
The superflavor symmetry is a symmetry between a heavy antiquark, $\bar{Q}$, and a heavy diquark, $QQ$, both of which carry the same color representation $\bar{\mathbf{3}}_c$. Thus, one obtains the superflavor partner structure by replacing $\bar{Q}$ with $QQ$ in heavy hadrons.
For instance, this symmetry establishes a partner structure between $\bar{D}(\bar{c}q)$ and $\Xi_{cc}(ccq)$. 
Since light mesons are coupled to the light quarks in the heavy hadrons, the superflavor symmetry allows us to use the same parameters for the interaction Lagrangians of $\bar{D}^{(*)}$ mesons and $\Xi_{cc}^{(*)}$ baryons.

One of the attractive exotic hadrons is the doubly charmed tetraquark $T_{cc}$ reported by the LHCb experiment, and this state is considered to be an isoscalar state with spin-parity $J^P = 1^+$~\cite{LHCb:2021vvq,LHCb:2021auc}. 
The quark content of $T_{cc}$ is $cc\bar{u}\bar{d}$, and this state is a genuine exotic hadron because its quantum number, namely charm number $2$ and baryon number $0$, cannot be reached in the ordinary meson and baryon pictures. 
Moreover, $T_{cc}$ lies very close to the $D^0 D^{\ast +}$ threshold. 
In fact, the mass difference $\delta m$ measured from the corresponding threshold, is obtained as $\delta m = -273\,\mathrm{keV}$ from the Breit--Wigner amplitude~\cite{LHCb:2021vvq}, while the pole analysis gives $\delta m = -360\,\mathrm{keV}$~\cite{LHCb:2021auc}. 
{This observation implies that this state is a $DD^\ast$ molecule.}
Consequently, a lot of studies based on the hadronic molecular model have been performed using the effective model in Refs.~\cite{Tornqvist:1993ng,Ohkoda:2012hv,Li:2012ss,Chen:2022asf,Wang:2021ajy,Wang:2021yld,Ren:2021dsi,Asanuma:2023atv,Sakai:2023syt,Sakai:2025djx} and the Lattice QCD in Refs.~\cite{Ikeda:2013vwa,Padmanath:2022cvl,Lyu:2023xro,Nagatsuka:2025szy}. 
We have studied $T_{cc}$ as a $D^{(\ast)}D^{(\ast)}$ molecule within the coupled channel framework, motivated by the small mass differences between $D$ and $D^\ast$ mesons~in Ref.~\cite{Sakai:2023syt}. In our analysis, we employed the one boson exchange potential (OBEP), where the $\pi$, $\rho$, $\omega$ and $\sigma$ mesons are exchanged. 

As mentioned before, the doubly charmed baryons $\Xi_{cc}^{(\ast)}$ are superflavor partners of the charmed mesons $\bar{D}^{(\ast)}$. Accordingly, in this study we consider the meson-baryon molecule $\barDs\Xis$ and baryon-baryon molecule $\Xis\Xis$ as a superflavor partner of $T_{cc}${, which were investigated by the nonrelativistic quark model in Ref.~\cite{Lu:2026iwm} and hadron molecule pictures in Refs.~\cite{Asanuma:2023atv,Yang:2019rgw}}. 
Analogous to the analysis of $T_{cc}$, we employ the OBEP to describe the interactions in the $\barDs\Xis$ and $\Xis\Xis$ molecules, and predict bound and resonant states. Furthermore, we use  the same set of parameters such as coupling constants and a cutoff parameter as those adopted in the $T_{cc}$ analysis as dictated by the superflavor symmetry. Here, we also consider the dependence of a sigma coupling constant.

This paper is organized as follows. In Sec.~\ref{sec;formalism}, we introduce the formalism of the OBEP. 
In Sec.~\ref{sec;Numerical}, we show the bound and resonant states of $\barDs\Xis$ and $\Xis\Xis$. We also discuss the mass spectra of $\barDs\Xis$ and $\Xis\Xis$ depending on the coupling constant of a $\sigma$ meson. 
Finally, in Sec.~\ref{sec;summary}, we summarize our results and discussions.

\section{Formalism}\label{sec;formalism}

In this study, we analyze bound and resonant states of $\barDs\Xis$ and $\Xis\Xis$ based on the OBEP under the HQS and superflavor symmetry. 
In order to construct the Lagrangians between a heavy meson and a light boson, we introduce the anti-heavy meson field $H^a$, which is a linear combination of the anti-heavy pseudoscalar meson field $\bar{D}$ and anti-heavy vector meson field $\bar{D}^\ast$~\cite{Grinstein:1992qt,Asanuma:2023atv}: 
\begin{equation}
    H^a = \left[ 
        \bar{D}^{a \ast}_\mu \gamma^\mu - \bar{D}^a \gamma_5     
    \right]\frac{1 - \Slash{v}}{2},
\end{equation}
where $a$ is an isospin index and $v$ is the four velocity with $v^0 > 0$ and $v^2 = 1$. 
This conjugate field is defined as 
\begin{equation}
    \bar{H}_a = \gamma^0 H^{a\dagger}\gamma^0 =  \frac{1 - \Slash{v}}{2}\left[ 
        \bar{D}^{\ast \dagger}_{a \mu}\gamma^\mu - \bar{D}^\dagger_a\gamma_5
     \right].
\end{equation}

Using these fields, we construct the effective Lagrangians between the heavy meson and the pion, vector mesons ($\rho$, $\omega$) and $\sigma$ mesons, which are invariant under the $SU(2)$ flavor and heavy quark spin symmetries~\cite{Asanuma:2023atv,Yamaguchi:2022oqz}: 
\begin{widetext}
    \begin{align}
        \La_{H}^\pi 
        =& ig\tr\left[ \bar{H}_{{a}} \gamma_\mu \gamma_5 A^{\mu}_{{ab}}H_{{b}} \right]\\
        =&
        \frac{g}{f_\pi}(\bar{D}^{\ast\mu\dagger}_a \bar{D}_b + \bar{D}^\dagger_a \bar{D}^{\ast\mu}_b)\partial_\mu(\vec{\pi}\cdot \vec{\tau})_{ba} 
        -i\frac{g}{f_\pi}\epsilon^{\mu\nu\rho\sigma}v_\mu \bar{D}^{\ast\dagger}_{a\nu}\bar{D}^\ast_{b\rho}\partial_\sigma(\vec{\pi}\cdot\vec{\tau})_{ba},\\
        \La_{H}^v 
    =&
    i\beta \tr\left[ \bar{H}_{{a}} v^\mu (\rho_\mu)_{{ab}}H_{{b}} \right] + i\lambda\tr\left[ 
        \bar{H}_{{a}}\sigma^{\mu\nu}F_{\mu\nu}(\rho)_{{ab}}H_{{b}} 
      \right]\nonumber\\
    =& \sqrt{2}\beta g_V \bar{D}_{{b}}\bar{D}_{{{a}}}^{\dagger} {{v \cdot(\hat{\rho})_{ba}} +}
      2\sqrt{2}\lambda g_V \epsilon_{{\mu\nu\alpha\beta}}(\bar{D}^{\ast{{\nu}}\dagger}_{{a}}\bar{D}_{{b}} + \bar{D}^\dagger_{{a}} \bar{D}^{\ast{\nu}}_{{b}})\partial^{{\alpha}}(\hat{\rho}^{{\beta}})_{{{ba}}}\\
    &- \sqrt{2}\beta g_V \bar{D}^{\ast\dagger}_{\mu{a}} v\cdot(\hat{\rho})_{ba}\bar{D}^{\ast\mu}_{{b}} + 2\sqrt{2}i\lambda g_V \bar{D}^\ast_{\mu{b}} \bar{D}^{\ast\dagger}_{\nu{a}}(\partial^\mu(\hat{\rho}^\nu)_{{ba}} - \partial^\nu(\hat{\rho}^\mu)_{{ba}}),\\
        \La^\sigma_H
        =&
        g_\sigma \tr\left[ \bar{H}\sigma H \right]\nonumber\\
        =
        &2g_\sigma \bar{D}^{\ast\dagger}\cdot \bar{D}^\ast \sigma {-} 2g_\sigma \bar{D}^\dagger \bar{D}\sigma.
    \end{align}
\end{widetext}
Here, $f_\pi=93\,\mathrm{MeV}$ denotes the pion decay constant. The axial current $A^\mu$ and pseudoscalar field $\vec{\pi}$ are defined as 
\begin{align}
    A^\mu = \frac{1}{2}&(\xi^\dagger \partial^\mu \xi - \xi\partial^\mu \xi^\dagger), \qquad \xi = \exp\left( i\frac{\vec{\pi}\cdot\vec{\tau}}{2f_\pi} \right),\\
    \vec{\pi}\cdot\vec{\tau} &= \begin{pmatrix}
        \pi^0 & \sqrt{2}\pi^+ \\
        \sqrt{2}\pi^- & - \pi^0
    \end{pmatrix}.
\end{align}
The field strength tensor and vector field are also defined as 
\begin{align}
    &F_{\mu\nu}(\rho) = \partial_\mu \rho_\nu - \partial_\nu \rho_\mu - [\rho_\mu, \rho_\nu],\\
    &\rho_\mu = i\frac{g_V}{\sqrt{2}}\hat{\rho}_\mu,\\
    &\hat{\rho}_\mu = \frac{1}{\sqrt{2}}\begin{pmatrix}
        \rho^0 + \omega & \sqrt{2}\rho^+ \\
        \sqrt{2}\rho^- & -\rho^0 + \omega
    \end{pmatrix}_\mu = \frac{1}{\sqrt{2}}(\vec{\tau}\cdot\vec{\rho}_\mu + \omega_\mu \mathbf{1}),
\end{align}
where $g_V = m_\rho/\sqrt{2}f_\pi$ is the coupling constant of a vector meson.

We also introduce the superfield for the heavy baryon, $\psi^\mu$, written as a linear combination of the doubly heavy baryon fields $\B(\frac{1}{2}^+)$ and $\B^{\ast \mu}( \frac{3}{2}^+ )$~\cite{Asanuma:2023atv}:
\begin{equation}
    \psi^\mu = \B^{\ast\mu} + \sqrt{\frac{1}{3}}(\gamma^\mu + v^\mu)\gamma_5 \B. 
\end{equation}
Its conjugate field $\bar{\psi}^\mu$ is given by 
\begin{equation}
    \bar{\psi}^\mu = \bar{\B}^{\ast\mu} - \sqrt{\frac{1}{3}}\bar{\B}\gamma_5 (\gamma^\mu + v^\mu).
\end{equation}
Here, $\B$ and $\B^\ast$ donote the Dirac spinor and Rarita-Schwinger spinor, which satisfy the constraints $\B^{(\ast)} = \frac{1 + \Slash{v}}{2}\B^{(\ast)}$, respectively.

Under the superflavor symmetry, the effective interaction Lagrangians for the doubly heavy baryon with the pion, vector mesons, and the $\sigma$ meson are obtained as:
\begin{widetext}
    \begin{align}
    \La^\pi_\psi 
    =& 
    ig \bar{\psi}^\mu \gamma_\nu \gamma_5 A^\nu \psi_\mu\nonumber\\
    =&
    - \frac{g}{2f_\pi}\bar{\B}^{\ast\mu}\partial_\nu(\vec{\pi}\cdot\vec{\tau})\gamma^\nu \gamma_5\B^\ast_\mu+ \sqrt{\frac{1}{3}}\frac{g}{f_\pi}\left( \bar{\B}^{\ast\mu}\partial_\mu(\vec{\pi}\cdot\vec{\tau})\B + \bar{\B}\partial_\mu(\vec{\pi}\cdot\vec{\tau})\B^{\ast\mu} \right)- \frac{g}{6f_\pi}\bar{\B}\partial_\nu(\vec{\pi}\cdot\vec{\tau})\gamma^\nu \gamma_5 \B\\
    \La^v_\psi 
    =& i\beta\bar{\psi}^\mu v\cdot \rho \psi_\mu + i \lambda \bar{\psi}^\mu \sigma^{\alpha\beta}F_{\alpha\beta}(\rho)\psi_\mu\nonumber\\
    =& -\frac{{\beta}g_V}{\sqrt{2}}\left( \bar{\B}^{\ast\mu}v\cdot\hat{\rho}\B^\ast_{\mu} - \bar{\B}v\cdot\hat{\rho}\B \right)-i\frac{{\lambda}g_V}{\sqrt{2}}\bar{\B}^{\ast\mu} \gamma_\alpha \gamma_\beta (\partial^\alpha \hat{\rho}^\beta - \partial^\beta \hat{\rho}^\alpha) \B^\ast_{\mu}\nonumber\\
    &+i2\sqrt{\frac{2}{3}}{{\lambda}}\bar{\B}^\ast_{\alpha}\gamma_\beta \gamma_5 (\partial^\alpha \hat{\rho}^\beta - \partial^\beta \hat{\rho}^\alpha)\B +i2\sqrt{\frac{2}{3}}{\lambda}\bar{\B} \gamma_\alpha \gamma_5 (\partial^\alpha \hat{\rho}^\beta - \partial^\beta \hat{\rho}^\alpha)\B^\ast_{ \beta}
    -i\frac{{\lambda}}{3\sqrt{2}}\bar{\B}\gamma_\alpha \gamma_\beta (\partial^\alpha \hat{\rho}^\beta - \partial^\beta \hat{\rho}^\alpha)\B\label{eq;Lvpsi_ex}\\
    \La^\sigma_\psi =& g_\sigma\bar{\psi}^\mu \sigma \psi_\mu\nonumber\\
    =&
    {-}{g}_\sigma \bar{\B}_{QQ}\sigma\B_{QQ} {+} {g}_\sigma \bar{\B}^{\ast\mu}_{QQ}\sigma\B^\ast_{QQ\mu}. \label{eq;Lsigmapsi_ex}
\end{align}
\end{widetext}
We note that the coupling constants in $\La^{\mathrm{boson}}_\psi$ are the same with those in $\La^{\mathrm{boson}}_H$ owing to the superflavor symmetry.

The potential in the momentum space is expressed in terms of the scattering amplitude $i\mathcal{M}$ for the $t$-channel Feynman diagram as
\begin{equation}
    V(q) = i\frac{i\M}{\sqrt{\prod_i 2m_i \prod_f 2m_f}},
\end{equation}
where $m_{i(f)}$ are the masses of the initial (final) states. 
When the Fourier transformation is implemented, the dipole-type form factor is attached to each vertex in order to {take account for} the hadronic size:
\begin{equation}
    F(\vec{q};m_\mathrm{ex}) = \frac{\Lambda^2 - m_{\mathrm{ex}}^2}{\Lambda^2 + \vec{q}^{\,\,2}}, 
\end{equation}
where $\Lambda$ and $m_{\mathrm{ex}}$ are the cutoff parameter and the mass of the exchanged meson, respectively. $\vec{q}$ is a momentum transfer carried by the exchanged meson. 
Following these procedures, the OBEPs are obtained. In particular, the one pion exchange potentials are given as
    \begin{align}
        &V^\pi(r) \nonumber\\
        =& k_\pi\frac{1}{3}\left( \frac{g}{2f_\pi} \right)^2 \left[ \vec{\Op}_1\cdot\vec{\Op}_2C(r;m_\pi) 
        + S_{12}(\hat{r})T(r;m_\pi)\right]\vec{\tau}_1\cdot\vec{\tau}_2,
    \end{align}
where $k_\pi$ is a constant depending on the process, and $C(r;m_\pi)$ and $T(r;m_\pi)$ are the central and tensor potentials, respectively:
\begin{widetext}
    \begin{align}
        C(r;m_{\mathrm{ex}})
        &=
        \frac{m_{\mathrm{ex}}^2}{4\pi}\left[ 
            \frac{e^{-m_{\mathrm{ex}}}}{r}
            -
            \frac{e^{-\Lambda r}}{r}
            -\frac{\Lambda^2 - m_\ex^2}{2\Lambda}e^{-\Lambda r}
         \right],\\
        T(r;m_\ex)
        &=
        \frac{1}{4\pi}\left[ 
            (3 + 3m_\ex r + m_\ex^2 r^2)\frac{e^{-m_\ex r}}{r^3}
            -
            (3 + 3\Lambda r + \Lambda^2 r^2)\frac{e^{-\Lambda r}}{r^3}
            +
            \frac{m_\ex^2 - \Lambda^2}{2}(1 + \Lambda r)\frac{e^{-\Lambda r}}{r}
         \right].
    \end{align}
\end{widetext} 
We note that the energy transfer at the vertex is neglected, and the contact term, which represents the short range potential,  is removed because the one light-boson exchange potentials are the long or middle range potentials. 
In addition, $\vec{\OC}$ denotes the spin operators of the charmed meson or the doubly charmed baryon, and $S_{12}(\hat{r})$ is the tensor operator.

\renewcommand{\arraystretch}{1.25}
    \begin{table}[tb]
        \caption{Coupling constants of the effective Lagrangians~\cite{CLEO:2001foe,Liu:2019stu,Isola:2003fh,Li:2012ss,Bardeen:2003kt,Liu:2008xz}. 
        $g_{\sigma NN}$ is a coupling constant between a $\sigma$ meson and a nucleon, while $g_\pi$ is determined by the Goldberger-Treiman relation.}
        \centering
        \begin{tabular}{ccc} \toprule[0.3mm]
        Coupling constants     &  Values &\\ \midrule[0.1mm]
        $g$     &  $0.59$ & $D^\ast \to D\pi$ decay \\
        $\beta$  & $0.9$ & Lattice QCD\\
        $\lambda$ & $0.56\,\mathrm{GeV^{-1}}$ & $B$ meson decay\\
        $g_\sigma^L$ & $3.4$ & $g_{\sigma NN}/3$\\ 
        $g_\sigma^S$ & $0.76$ & ${g_\pi}/2\sqrt{6}$\\
        \bottomrule[0.3mm]
        \end{tabular}
        \label{tab;coupling_constant}
    \end{table}
\renewcommand{\arraystretch}{1.00}
Finally, the parameters are summarized in Table.~\ref{tab;coupling_constant}. 
The pion coupling constant $g$ is determined by the $D^\ast \to D\pi$ decay~\cite{CLEO:2001foe}, and $\beta$ and $\lambda$ are determined by the Lattice QCD and $B$ meson decay~\cite{Liu:2019stu,Isola:2003fh}, respectively. 
The coupling constant $g_\sigma$ is uncertain, but there are two typical values. 
{
The large $g_\sigma$, denoted by $g_\sigma^L$, is determined from $g_{\sigma NN}/3$, where $g_{\sigma NN}$ is the coupling constant between the $\sigma$ meson and the nucleon~\cite{Isola:2003fh}.
This large coupling constant is estimated based on quark-model counting. Since the one-boson exchange occurs at the light-quark level, the coupling constants for the $\barDs\Xis$ and $\Xis\Xis$ systems are related to $g_{\sigma NN}$ by a factor of $1/3$. Thus, this value is determined from the parameters of the nuclear force.}
{On the other hand, the small $g_\sigma$, denoted by $g_\sigma^S$, is determined as ${g_\pi}/(2\sqrt{6})$ by comparing the effective Lagrangian for the $\sigma$ meson~\cite{Liu:2008xz} with the linear sigma model Lagrangian~\cite{Bardeen:2003kt}.
Here, $g_\pi$ is the $0^+ \to 0^- \pi$ coupling constant determined from the Goldberger-Treiman relation and the observed mass difference between $D_s(0^+)$ and $D_s(0^-)$~\cite{Li:2012ss,Bardeen:2003kt,Liu:2008xz}.
}

We also summarize the hadron masses in table~\ref{tab;mass_hadron}. The isospin symmetry is assumed for the masses of the heavy meson $\bar{D}$ and pion $\pi$. Moreover, the mass of $\Xi_{cc}^\ast$, which has not yet been observed, is determined using the relation derived from the superflavor symmetry~\cite{Hu:2005gf}:
\begin{equation}
    m_{\Xi_{cc}^\ast} - m_{\Xi_{cc}} = \frac{3}{4}(m_{\bar{D}^\ast} - m_{\bar{D}}).
\end{equation} 
\renewcommand{\arraystretch}{1.25}
    \begin{table}[tb]
        \caption{Masses of heavy hadrons and light mesons~\cite{ParticleDataGroup:2024cfk}. The values are given in units of MeV. }
        \centering
        \begin{tabular}{cc|cc} \toprule[0.3mm]
        Heavy hadron     &  Values [MeV] & Light meson & Values [MeV]\\ \midrule[0.1mm]
        $\bar{D}$ & $1868\,\mathrm{MeV}$ & $\pi$ & $138\,\mathrm{MeV}$ \\
        $\bar{D}^\ast$ & $2009\,\mathrm{MeV}$ & $\rho$ & $770\,\mathrm{MeV}$ \\
        $\Xi_{cc}$ & $3621\,\mathrm{MeV}$ & $\omega$ & $782\,\mathrm{MeV}$ \\
        $\Xi_{cc}^\ast$ & $3727\,\mathrm{MeV}$ & $\sigma$ & $500\,\mathrm{MeV}$ \\
        \bottomrule[0.3mm]
        \end{tabular}
        \label{tab;mass_hadron}
    \end{table}
\renewcommand{\arraystretch}{1.00}

\section{Numerical calculation}\label{sec;Numerical}

In this section, we investigate the bound and resonant states of $\barDs\Xis$ and $\Xis\Xis$ with dependence on $g_\sigma$. 

First, the cutoff parameter $\Lambda$ is determined to reproduce the binding energy of $T_{cc}$, $360\,\mathrm{keV}$, for both cases of $g^L_\sigma$ and $g^S_\sigma$ by solving the coupled channel Schr\"{o}dinger equation{~\cite{Sakai:2023syt}} using the Gaussian expansion method~\cite{Hiyama:2003cu,Hiyama:2018ivm}. 
{We find that $\Lambda_L = 1075.9\,\mathrm{MeV}$ for $g_\sigma^L$ and $\Lambda_S= 1683.8\,\mathrm{MeV}$ for $g_\sigma^S$ reproduce the binding energy of $T_{cc}$}.
In the case of $g^L_\sigma$, the $\sigma$ exchange force plays a major role in binding $T_{cc}$, as discussed in Ref.~\cite{Asanuma:2023atv, Sakai:2023syt}. 
For case~$g^S_\sigma$, however, the contribution of the $\sigma$ exchange becomes weaker. 
As a result, a larger cutoff parameter is required for~$g^S_\sigma$ than for~$g^L_\sigma$, which indicates that the one $\pi$ and one $\rho$ exchange forces provide stronger contributions~\cite{Asanuma:2023atv}.

\subsection{$\barDs \Xis$ molecule \label{sec;DXi_molecule}}

\renewcommand{\arraystretch}{1.25}
    \begin{table*}[tb]
        \caption{Channels of $\barDs\Xis$ with spin-parity $J^P$~\cite{Asanuma:2023atv}. We use the notation $^{2S+1}L_J$ with $S$, $L$, and $J$ being the spin, angular momentum and total momentum, respectively. }
        \centering
        \begin{tabular}{cccc} \toprule[0.3mm]
        $J^P$     & Channels\\ \midrule[0.1mm]
        $\frac{1}{2}^-$ & $\bar{D}\Xi_{cc}(^2S_{1/2})$, $\bar{D}\Xi_{cc}^\ast(^4D_{1/2})$, $\bar{D}^\ast\Xi_{cc}(^2S_{1/2}, ^4D_{1/2})$, $\bar{D}^\ast\Xi_{cc}^\ast (^2S_{1/2}, ^4D_{1/2}, ^6D_{1/2})$\\
        $\frac{3}{2}^-$ & $\bar{D}\Xi_{cc}(^2D_{3/2})$, $\bar{D}\Xi_{cc}^\ast(^4S_{3/2}, ^4D_{3/2})$, $\bar{D}^\ast\Xi_{cc}(^2D_{3/2}, ^4S_{3/2}, ^4D_{3/2})$, $\bar{D}^\ast\Xi_{cc}^\ast (^2D_{3/2}, ^4S_{3/2}, ^4D_{3/2}. ^6D_{3/2}, ^6G_{3/2})$\\
        $\frac{5}{2}^-$ & $\bar{D}\Xi_{cc}(^2D_{5/2})$, $\bar{D}\Xi_{cc}^\ast(^4D_{5/2}, ^4G_{5/2})$, $\bar{D}^\ast\Xi_{cc}(^2D_{5/2}, ^4D_{5/2}, ^4G_{5/2})$, $\bar{D}^\ast\Xi_{cc}^\ast (^2D_{5/2}, ^4D_{5/2}, ^4G_{5/2}. ^6S_{5/2}, ^6D_{5/2}, ^6G_{5/2})$\\
        \bottomrule[0.3mm]
        \end{tabular}
        \label{tab;Channel_DXi}
    \end{table*}
\renewcommand{\arraystretch}{1.00}

\begin{figure*}[tb]
    \centering
    \includegraphics[width=0.8\linewidth]{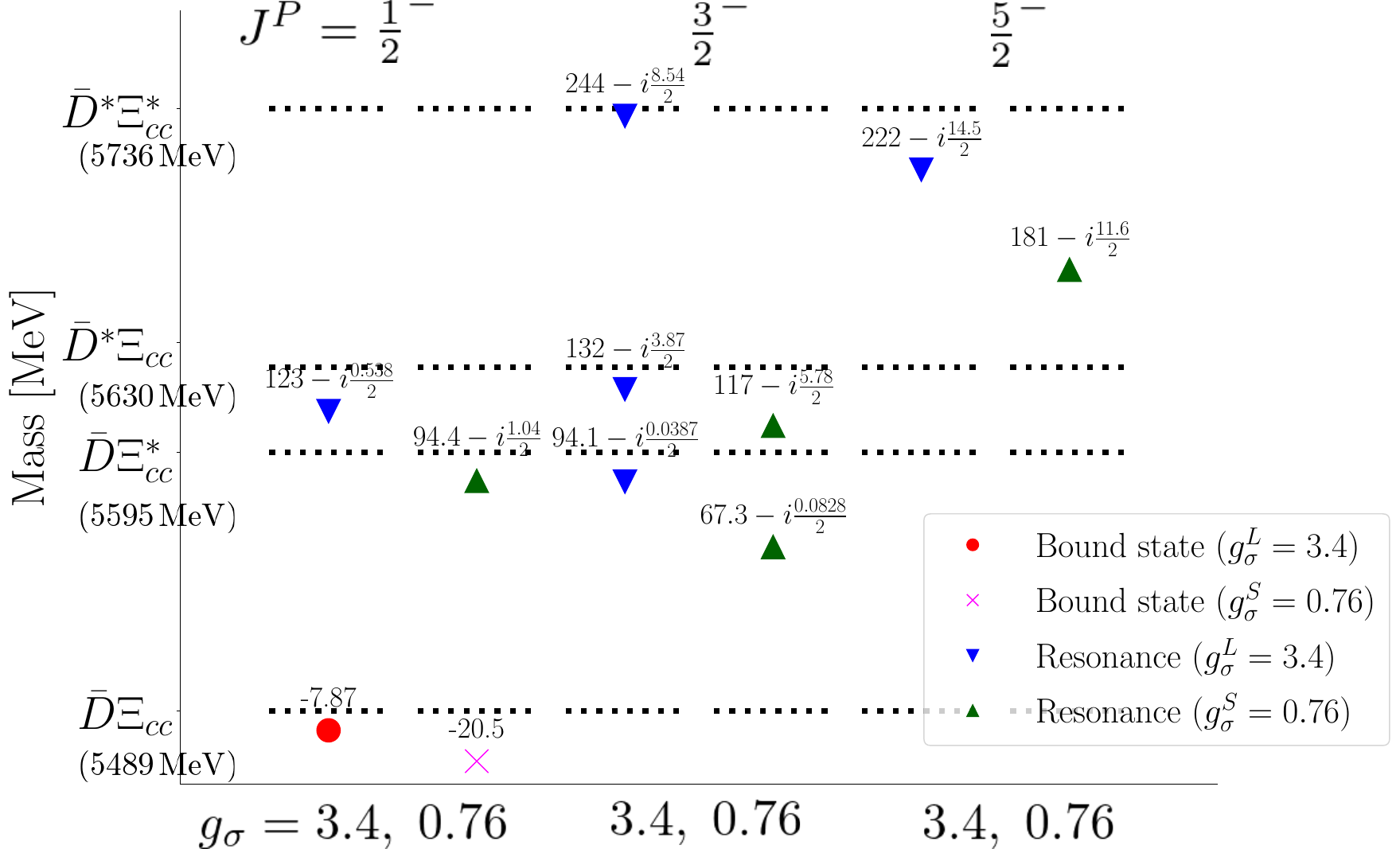}
    \caption{Masses of the isoscalar $\barDs\Xis$ with $\frac{1}{2}^-$, $\frac{3}{2}^-$ and $\frac{5}{2}^-$. The black dotted line shows the thresholds. The red circle and magenta cross show the masses of the bound states with $g_\sigma^L$ and $g_\sigma^S$, respectively. The blue upward triangles and green downward triangles show the masses of the resonances with $g_\sigma^L$ and $g_\sigma^S$, respectively. 
    Numerical values displayed in the figure correspond to the eigenvalues $-B$ for bound states and $E_r - i\frac{\Gamma}{2}$ for resonances. 
    The values are given in units of MeV. 
    }
    \label{fig;mass_DXi}
\end{figure*}

In this subsection, we show the numerical results for $\barDs\Xis$ with $J^P = \frac{1}{2}^-, \frac{3}{2}^-$ and $\frac{5}{2}^-$ for $I=0, 1$. 
We solve the coupled-channel Schr\"{o}dinger equation by the Gaussian expansion method and complex scaling method~\cite{Suzuki:2005wv,Myo:2014ypa,Myo:2020rni}. The channels for $\barDs\Xis$ with the corresponding spin-parity $J^P$ are listed in Table.~\ref{tab;Channel_DXi}.  
The mass spectra of the isoscalar $\barDs\Xis$ with $g_\sigma = g_\sigma^L$ and $g_\sigma = g_\sigma^S$ are depicted in Fig.~\ref{fig;mass_DXi}, while the numerical results for both the isoscalar and isovector $\barDs\Xis$ are summarized in Table~\ref{tab;mass_DXi}. These results indicate the appearance of bound and resonant states in the $\barDs\Xis$ system with $J^P = \frac{1}{2}^-$, $\frac{3}{2}^-$ and $\frac{5}{2}^-$. 

\renewcommand{\arraystretch}{1.25}
    \begin{table}[tb]
        \centering
        \caption{Eigenvalues of $\barDs\Xis$. The first column shows $I(J^P)$ with $I$, $J$ and $P$ with the isospin, spin and parity, respectively. 
        The third and fourth columns show the complex-scaled energy eigenvalues of the bound and resonant states for $g_\sigma^L$ and $g_\sigma^S$, respectively. Here, $-B$ and $E_r$ denote the masses measured from the lowest thresholds for bound states and resonances, respectively, and $\Gamma$ represents the decay width of the resonances. The values are given in units of MeV. 
        }\label{tab;mass_DXi}
        \begin{tabular}{cccc}
            \toprule[0.3mm]
            \multirow{2}{*}{$I(J^P)$} & \multirow{2}{*}{lowest threshold} & $g^L_\sigma = 3.4$ & $g^S_\sigma = 0.76$\\
            &&\multicolumn{2}{c}{$-B$, $E_r - i\frac{\Gamma}{2}$} \\
            \midrule[0.1mm]
            \multirow{2}{*}{$0(\frac{1}{2}^-)$} & \multirow{2}{*}{$\bar{D}\Xi_{cc}$} & $-7.87$ & $-20.5$\\
            &&$123 - \frac{0.538}{2}$  & $94.4 - i\frac{1.04}{2}$ \\
            \midrule[0.1mm]
            $1(\frac{1}{2}^-)$ & {$\bar{D}\Xi_{cc}$} & $246 - i \frac{0.679}{2}$ & - \\
            \midrule[0.1mm]
            \multirow{3}{*}{$0(\frac{3}{2}^-)$} & \multirow{3}{*}{$\bar{D}\Xi_{cc}$} & $94.1 - i \frac{0.0387}{2}$ & $67.3 - i \frac{0.0828}{2}$\\
            &&$132 - i\frac{3.87}{2}$  & $117 - i\frac{5.78}{2}$ \\
            &&$244 - i \frac{8.54}{2}$ & - \\
            \midrule[0.1mm]
            $0(\frac{5}{2}^-)$ & $\bar{D}\Xi_{cc}$ & $222 - i \frac{14.5}{2}$ & $181 - i \frac{11.6}{2}$ \\
            \bottomrule[0.3mm]
        \end{tabular}
    \end{table}
\renewcommand{\arraystretch}{1.00}

\begin{figure*}
    \begin{tabular}{cc}
        $g_\sigma = g^L_\sigma$ & $g_\sigma = g^S_\sigma$ \\
        \includegraphics[width=0.5\linewidth]{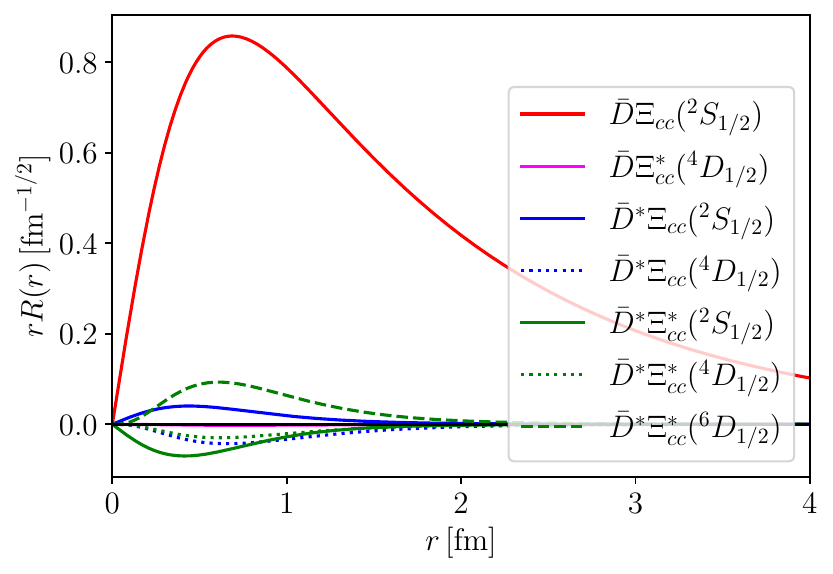}&
        \includegraphics[width=0.5\linewidth]{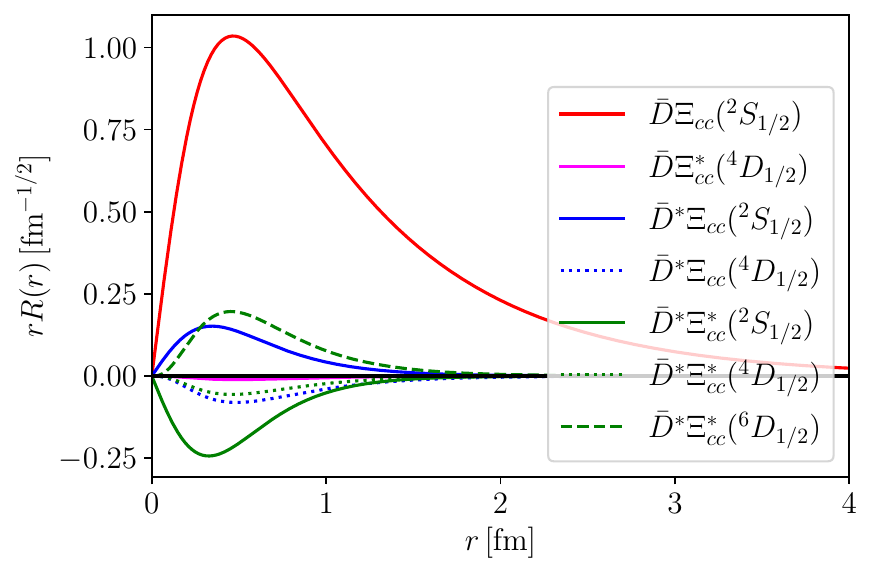}
    \end{tabular}
    \caption{Wavefunctions of $\barDs\Xis$ with $I(J^P) = 0(\frac{1}{2}^-)$. The left panel shows the wavefunction of $\barDs\Xis$ with $g_\sigma^L$, while right one shows the wavefunction of $\barDs\Xis$ with $g_\sigma^S$. The red, magenta, blue and green lines show the wavefunctions for $\bar{D}\Xi_{cc}$, $\bar{D}\Xi_{cc}^\ast$, $\bar{D}^\ast\Xi_{cc}$ and $\bar{D}^\ast\Xi_{cc}^\ast$. }
    \label{fig;wavefunction_DXi}
\end{figure*}

\begin{figure}[tbp]
    \centering
    \begin{center}
        \includegraphics[width=0.8\linewidth]{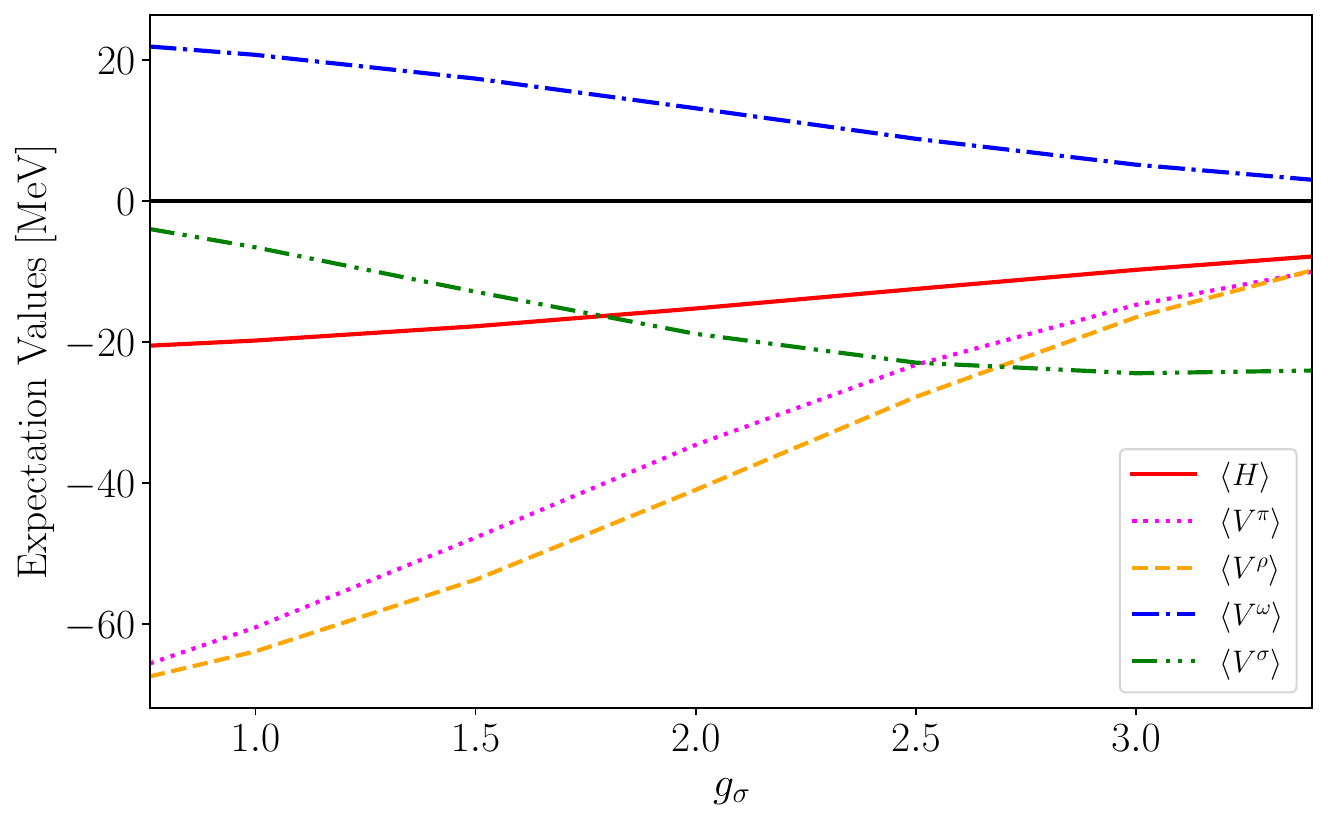}
    \end{center}
    \caption{$g_\sigma$ dependence of the expectation values of the Hamiltonian and OBEP for $\barDs\Xis$ with $0(\frac{1}{2}^-)$. The solid line shows the expectation value of the Hamiltonian $\braket{H}$, which is equivalent to the binding energy of $\barDs\Xis$, while the other lines show the expectation values of the one boson exchange potentials denoted by $\braket{V^\pi}$, $\braket{V^\rho}$, $\braket{V^\omega}$ and $\braket{V^\sigma}$. The cutoff parameter is determined to reproduce the binding energy of $T_{cc}$ for each $g_\sigma$. }
    \label{fig;DXi_Expectation_gsigma}
\end{figure}

First, we find the bound states of $\barDs\Xis$ with $0(\frac{1}{2}^-)$ which were also discussed in Ref.~\cite{Asanuma:2023atv}. 
As shown in Table~\ref{tab;Channel_DXi}, the $S$-wave $\bar{D}\Xi_{cc}$ component is present only in the $J^P=\frac{1}{2}^-$ state. This fact indicates the absence of bound states with $J^P=\frac{3}{2}^-$ and $\frac{5}{2}^-$, because a strong attraction is required to produce a bound state for $L \neq 0$. 
The wavefunctions of $\barDs\Xis$ with $0(\frac{1}{2}^-)$ are shown in Fig.~\ref{fig;wavefunction_DXi}, and their binding energies, mixing ratios and root mean squared (r.m.s) distances are summarized in Table~\ref{tab;Properties_DXi}. 
The binding energy of $\barDs\Xis$ with $g_\sigma = g^S_\sigma$ is larger than that with $g_\sigma = g^L_\sigma$. 
In the both cases, the $\bar{D}\Xi_{cc}(^2S)$ is dominant to bind $\barDs\Xis$. However, as for the other channels, the composition of the mixing ratios differs between the two $g_\sigma$ cases.
In order to clarify these differences, we examine the $g_\sigma$ dependence of the expectation value of the Hamiltonian (which corresponds to the binding energy) as well as those of one boson exchange potential as shown in Fig.~\ref{fig;DXi_Expectation_gsigma}. 
Here, the cutoff parameter is tuned to reproduce the binding energy of $T_{cc}$ for each value of $g_\sigma$.
As $g_\sigma$ decreases, the expectation value of the one $\sigma$ exchange potential becomes weaker, whereas those of the one $\pi$ and one $\rho$ exchange potentials become larger, reflecting the increase of the cutoff parameter $\Lambda$.
Consequently, the effects of the tensor force from the one $\pi$ exchange potential and the central force from the one $\rho$ exchange potential are enhanced. This leads to a larger binding energy for the $\barDs \Xis$ system with $g_\sigma = g_\sigma^S$ than for that with $g_\sigma = g_\sigma^L$, and enhances the mixing between the $\bar{D}\Xi_{cc}(^2S)$ channel and the other channels.

\renewcommand{\arraystretch}{1.25}
\begin{table}[tb]
    \centering
    \caption{The binding energy $B$, mixing ratios and root mean squared (r.m.s) distances of $\barDs\Xis$ with $0(\frac{1}{2}^-)$ for each case of $g_\sigma = g^L_\sigma$ and $g_\sigma = g^S_\sigma$. 
    The second and third columns show the properties in the case of $g_\sigma = 3.4$ and $g_\sigma = 0.76$, respectively. 
    The values of the binding energy, mixing ratio, and r.m.s distance are given in units of MeV, $\%$, and fm, respectively. }
    \begin{tabular}{ccc}
        \toprule[0.3mm]
        \multirow{2}{*}{Properties}& \multicolumn{2}{c}{$g_\sigma$}\\
        & $g^L_\sigma = 3.4$ & $g^S_\sigma = 0.76$\\
        \midrule[0.1mm]
        $B$ & $7.87\,\mathrm{MeV}$ & $20.5\,\mathrm{MeV}$ \\
        $\bar{D}\Xi_{cc}(^2S)$ & ${{98.8\,\%}}$ & $93.5\,\%$\\
        $\bar{D}\Xi^\ast_{cc}(^4D)$ & $0\,\%$ & $0\,\%$ \\
        $\bar{D}^\ast\Xi_{cc}(^2S)$ & $0.104\,\%$ & $1.18\,\%$\\
        $\bar{D}^\ast\Xi_{cc}(^4D)$ & $0.136\,\%$ & $0.378\,\%$\\
        $\bar{D}^\ast\Xi^\ast_{cc}(^2S)$ & $0.287\,\%$ & $2.78\,\%$\\
        $\bar{D}^\ast\Xi^\ast_{cc}(^4D)$ & $0.0603\,\%$ & $0.169\,\%$\\
        $\bar{D}^\ast\Xi^\ast_{cc}(^6D)$ & $0.586\,\%$ & $2.02\,\%$ \\
        r.m.s distance & $1.39$ fm  & $0.884$ fm\\
        \bottomrule[0.3mm]
    \end{tabular}\label{tab;Properties_DXi}
\end{table}
\renewcommand{\arraystretch}{1.00}

\begin{figure*}[tbp]
    \begin{tabular}{cc}
        $g_\sigma = g^L_\sigma$ & $g_\sigma = g^S_\sigma$ \\
        \includegraphics[width=0.5\linewidth]{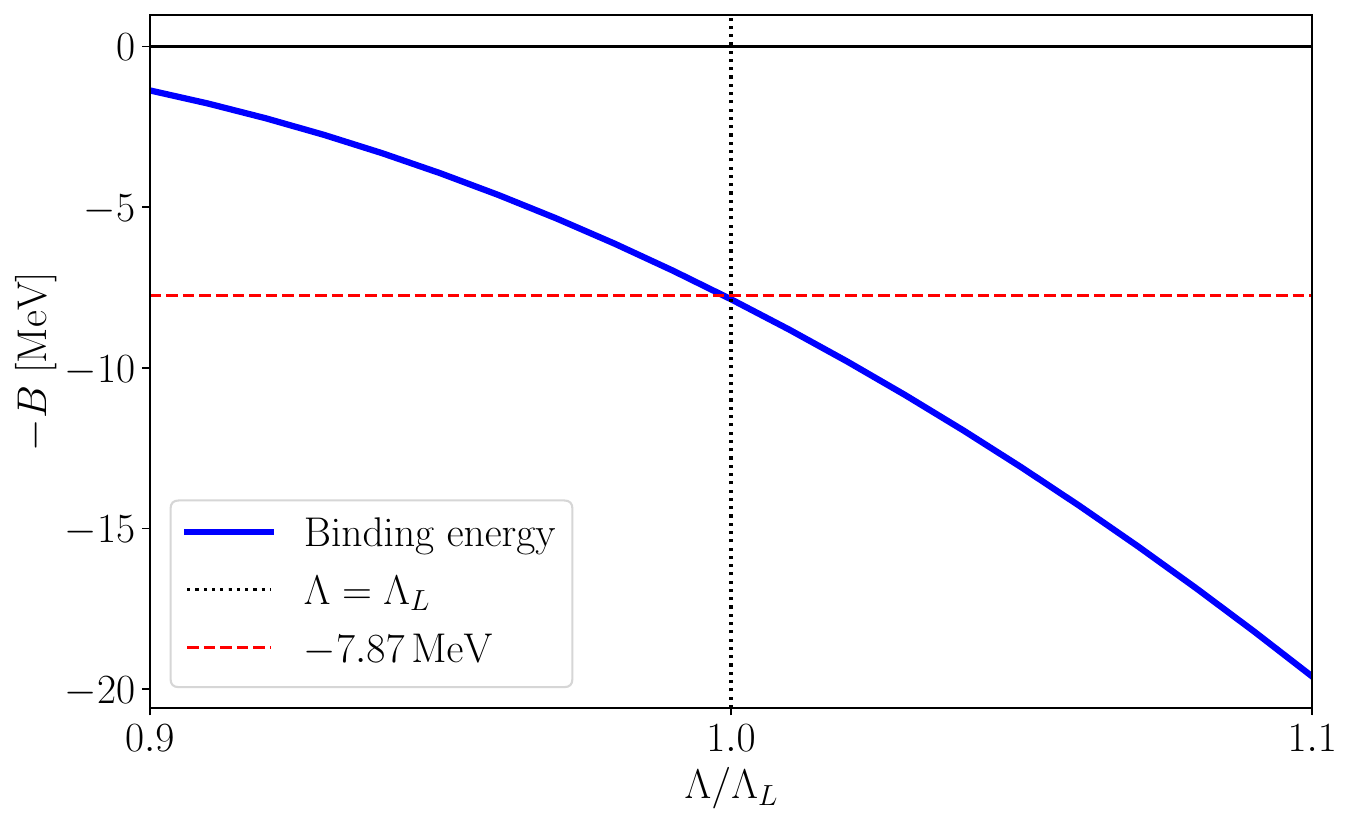}&
        \includegraphics[width=0.5\linewidth]{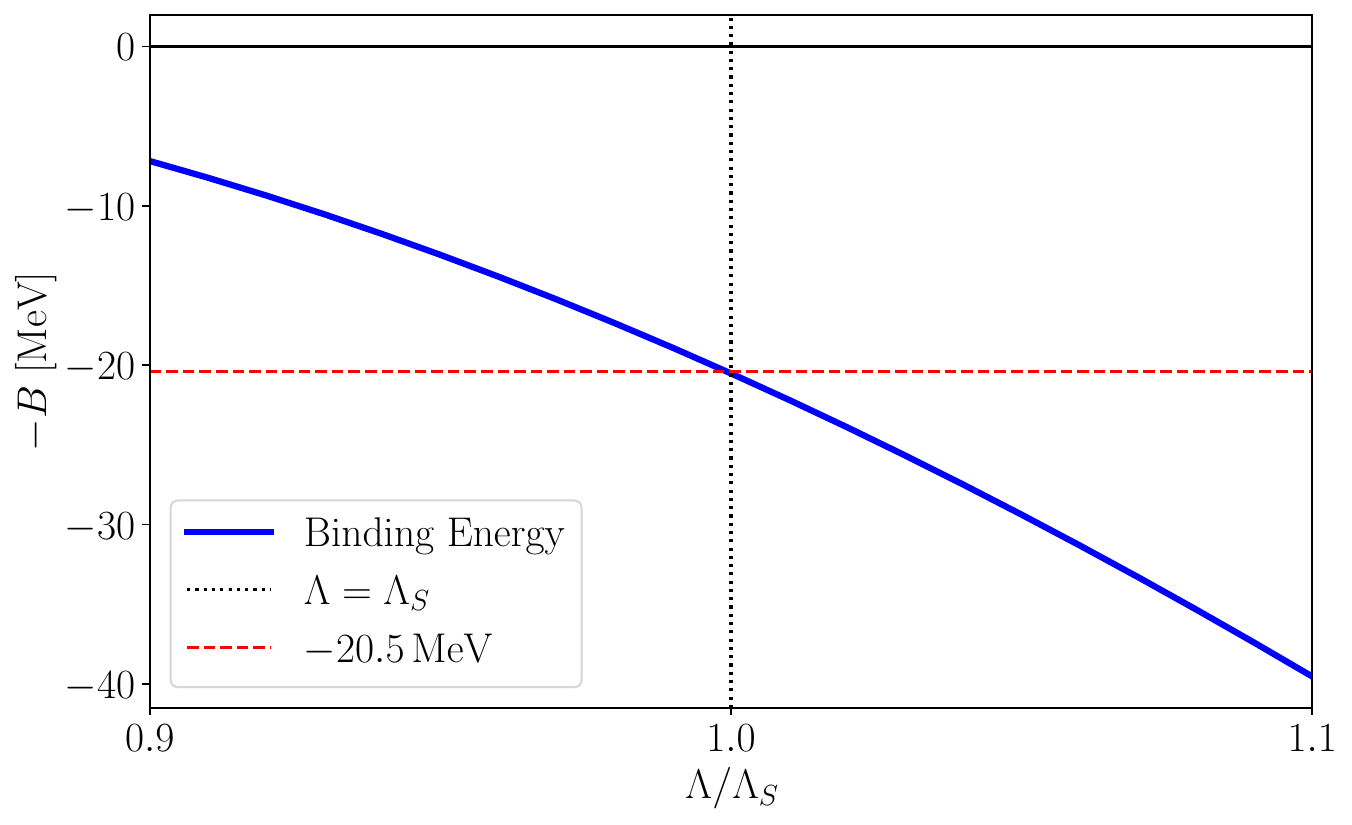}
    \end{tabular}
    \caption{Cutoff dependence of the binding energy of $\barDs\Xis$. The left panel shows the results for $g_\sigma^L$, and the right panel for $g_\sigma^S$. Here, $B$ denotes the binding energy of $\barDs\Xis$.}
    \label{fig;Lambda_dep_DXi}
\end{figure*}

In this study, we assume that the same cutoff parameter can be used for $\barDs$ and $\Xis$ owing to the superflavor symmetry.    Since the charm quark is not sufficiently heavy, HQS and superflavor symmetry breaking effects may affect the cutoff parameters through the hadron size dependence.
We show the cutoff parameter dependence of the binding energy of $\barDs\Xis$ with both $g_\sigma$. 
This result is depicted in Fig.~\ref{fig;Lambda_dep_DXi}, which indicates that the binding energies of $\barDs\Xis$ for both $g_\sigma^L$ and $g_\sigma^S$ increase as the cutoff parameter increases.

It is instructive to compare the present analysis with previous study employing similar formalisms. 
In Ref.~\cite{Asanuma:2023atv}, bound states of $\barDs\Xis$ were investigated within essentially the same formalism, except that the energy transfer at the vertex was taken into account.
Nevertheless, our results are consistent with those of Ref.~\cite{Asanuma:2023atv}, indicating that the contributions from the energy transfer are small.

Next, we discuss the resonances of $\barDs\Xis$ with $I = 0$. Resonances of $\barDs\Xis$ are found not only for $J^P = \frac{1}{2}^-$ but also for $J^P = \frac{3}{2}^-$ and $\frac{5}{2}^-$. 
In higher-spin states, the larger number of $D^\ast \Xi_{cc}^\ast$ channels strengthens the coupled-channel dynamics, thereby promoting the emergence of resonant states. 
These states are indicated as Feshbach resonances because $S$-wave bound states of $\barDs\Xis$ appear when the lower channels are cut. 
We obtain the resonance with $\frac{1}{2}^-$ below the $\bar{D}\Xi_{cc}^\ast$ threshold for $g_\sigma = g_\sigma^S$, while this channel has no $S$-wave component. In fact, we find that the main component of the resonance is the $\bar{D}^\ast \Xi_{cc}(^2S)$ channel, and hence, it is originated from the $\bar{D}^\ast \Xi_{cc}(^2S)$ bound state. For $\frac{1}{2}^-$, we find no resonance below the $\bar{D}^\ast \Xi_{cc}^\ast$ threshold, while there is $\bar{D}^\ast \Xi_{cc}^\ast(^2S)$ channel. Actually, for $g_\sigma = g_\sigma^L$, a bound state appears when the lower channels are cut. However, the bound state disappears due to the level repulsion caused by coupling to the lower channels. 
Except for the quasi bound state of $\bar{D}^\ast\Xi_{cc}^\ast$ with $\frac{3}{2}^-$, the quasi bound states with $g_\sigma = g^S_\sigma$ are bound more deeply than those with $g_\sigma = g^L_\sigma$ for the same reason as in the case of the bound state.  
The absence of a quasi bound state of $\bar{D}^\ast\Xi_{cc}^\ast$ for $J^P = \frac{3}{2}^-$ with $g_\sigma = g_\sigma^S$ is attributed to the fact that the one $\pi$ and one $\rho$ exchange potentials do not provide sufficient attraction.
Indeed, no bound state of $\bar{D}^\ast\Xi_{cc}^\ast$ with $J^P = \frac{3}{2}^-$ appears when the $\bar{D}\Xi_{cc}$, $\bar{D}\Xi_{cc}^\ast$, and $\bar{D}^\ast\Xi_{cc}$ channels are removed.
In the similar way, a bound state with $\frac{1}{2}^-$ below the $\bar{D}^\ast \Xi_{cc}^\ast$ threshold is not found even if the lower channels are ignored.

Finally, we consider the isovector $\bar{D}^{(*)}\Xi_{cc}^{(*)}$ state. 
We find a resonant state only for $I(J^P)=1(\frac{1}{2}^-)$ with $g_\sigma=g_\sigma^L$, with a complex-scaled energy of $246 - i \frac{0.679}{2}\,\mathrm{MeV}$. 
The state is primarily bound by the one $\sigma$ exchange potential. As discussed above, in the $g_\sigma^S$ case, $\barDs\Xis$ with $I=0$ is bound mainly by the one $\pi$ and one $\rho$ exchange potentials. In contrast, for $I=1$, the one $\pi$ and one $\rho$ exchange potentials provide weaker attractions because of the isospin factor $\vec{\tau}_1\cdot\vec{\tau}_2$. 
Consequently, no $\barDs\Xis$ state with $I = 1$ is obtained for $g_\sigma = g_\sigma^S$.

\subsection{$\Xis \Xis$ molecule \label{sec;XiXi_molecule}}

\renewcommand{\arraystretch}{1.25}
    \begin{table*}[tb]
        \caption{Channels of $\Xis\Xis$ with some quantum numbers $I(J^P)$ where $I$ is the isospin. We use $^{2S+1}L_J$ as used in Table~\ref{tab;Channel_DXi}, and $[\Xi_{cc}\Xi_{cc}^\ast]_\pm = \frac{1}{\sqrt{2}}(\Xi_{cc}\Xi_{cc}^\ast \pm \Xi_{cc}^\ast \Xi_{cc})$. }
        \centering
        \begin{tabular}{cccc} \toprule[0.3mm]
        $I(J^P)$     & Channels\\ \midrule[0.1mm]
        $0(0^+)$ & $[\Xi_{cc}\Xi_{cc}^\ast]_+ (^5D_0)$ \\
        $1(0^+)$ & $\Xi_{cc}\Xi_{cc}(^1S_0)$, $[\Xi_{cc}\Xi_{cc}^\ast]_- (^5D_0)$, $\Xi_{cc}^\ast\Xi_{cc}^\ast(^1S_0, ^5D_0)$ \\
        $0(1^+)$ & $\Xi_{cc}\Xi_{cc}(^3S_1, ^3D_1)$, $[\Xi_{cc}\Xi_{cc}^\ast]_-(^3S_1, ^3D_1)$, $[\Xi_{cc}\Xi^\ast_{cc}]_+(^5D_1)$, $\Xi^\ast_{cc}\Xi^\ast_{cc}(^3S_1, ^3D_1, ^7D_1, ^7G_1)$\\
        $1(1^+)$ & $[\Xi_{cc}\Xi_{cc}^\ast]_+(^3S_1, ^3D_1)$, $[\Xi_{cc}\Xi^\ast_{cc}]_-(^5D_1)$, $\Xi^\ast_{cc}\Xi^\ast_{cc}(^5D_1)$\\
        $0(2^+)$ & $\Xi_{cc}\Xi_{cc}(^3D_2)$, $[\Xi_{cc}\Xi_{cc}^\ast]_-(^3D_2)$, $[\Xi_{cc}\Xi^\ast_{cc}]_+(^5S_2, ^5D_2, ^5G_2)$, $\Xi^\ast_{cc}\Xi^\ast_{cc}(^3D_2, ^7D_2, ^7G_2)$\\
        $1(2^+)$ & $\Xi_{cc}\Xi_{cc}(^1D_2)$, $[\Xi_{cc}\Xi_{cc}^\ast]_+(^3D_2)$, $[\Xi_{cc}\Xi^\ast_{cc}]_-(^5S_2, ^5D_2, ^5G_2)$, $\Xi^\ast_{cc}\Xi^\ast_{cc}(^1D_2, ^5S_2, ^5D_2, ^5G_2)$\\
        \bottomrule[0.3mm]
        \end{tabular}
        \label{tab;Channel_XiXi}
    \end{table*}
\renewcommand{\arraystretch}{1.00}

\begin{figure*}[tb]
    \centering
    \includegraphics[width=0.8\linewidth]{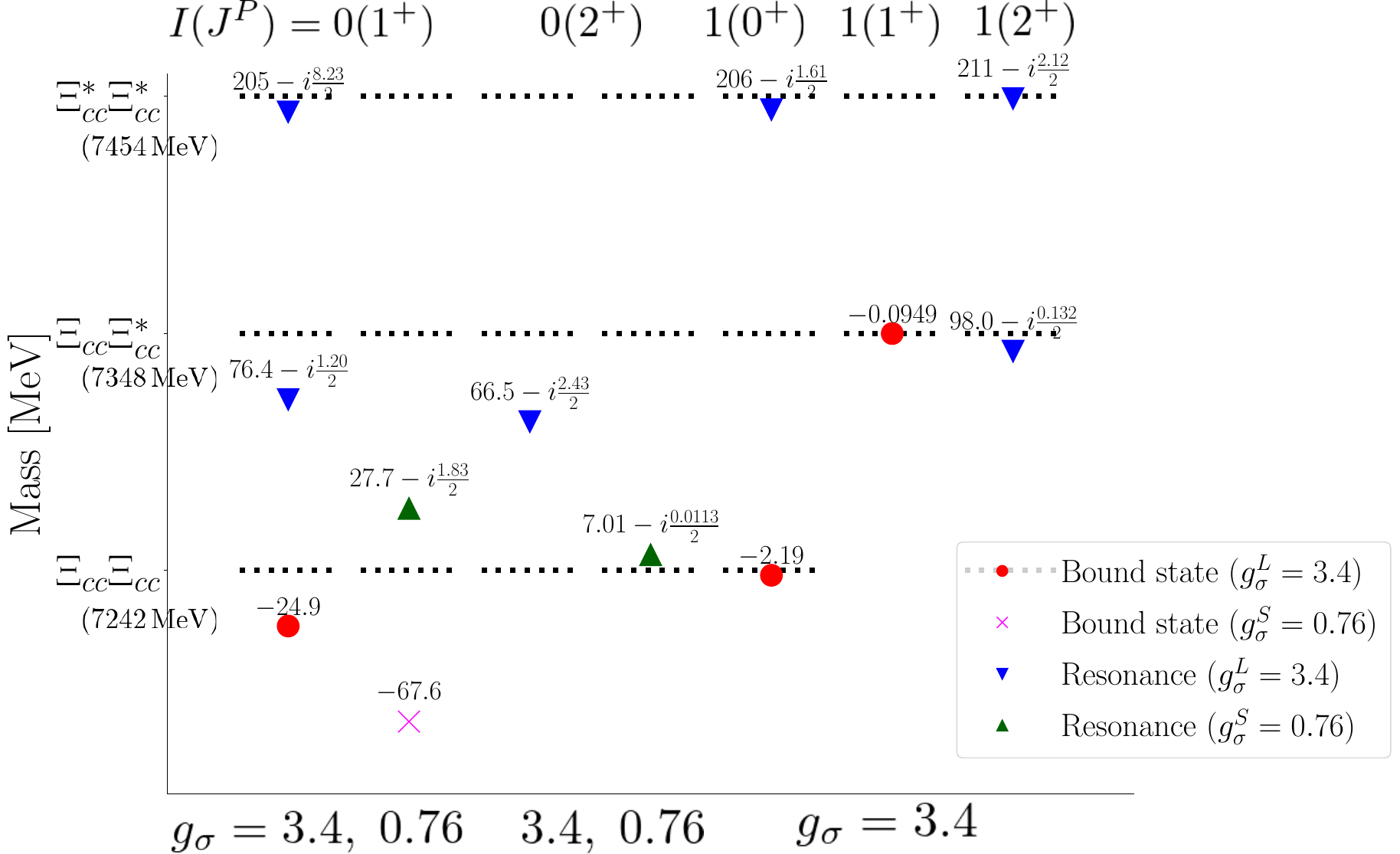}
    \caption{Masses of $\Xis\Xis$ with $0(1^+)$, $0(2^+)$, $1(0^+)$, $1(1^+)$ and $1(2^+)$. The same notation as Fig.~\ref{fig;mass_DXi} is used. }
    \label{fig;mass_XiXi}
\end{figure*}

\renewcommand{\arraystretch}{1.25}
\begin{table}[tb]
    \centering
    \caption{Eigenvalues of $\Xis\Xis$. The same conventions as Table~\ref{tab;mass_DXi} are used. }
    \begin{tabular}{cccc}
        \toprule[0.3mm]
        \multirow{2}{*}{$I(J^P)$} & \multirow{2}{*}{lowest threshold} & $g^L_\sigma = 3.4$ & $g^S_\sigma = 0.76$\\
            &&\multicolumn{2}{c}{$-B$, $E_r - i\frac{\Gamma}{2}$} \\
        \midrule[0.1mm]
        $0(0^+)$ & $\Xi_{cc}\Xi^\ast_{cc}$ & - & - \\        \midrule[0.05mm]
        \multirow{2}{*}{$1(0^+)$} & \multirow{2}{*}{$\Xi_{cc}\Xi_{cc}$} & $-2.19 $ & \multirow{2}{*}{-} \\
        &&  $206 - i\frac{1.61}{2}$ & \\  \midrule[0.05mm]
        \multirow{3}{*}{$0(1^+)$} & \multirow{3}{*}{$\Xi_{cc}\Xi_{cc}$} & $-24.9$ & $-67.6$ \\
        && $76.4 - i \frac{1.20}{2}$ & $27.7 - i\frac{1.83}{2}$ \\
        && $205 - i\frac{8.23}{2}$ & - \\ \midrule[0.05mm]
        $1(1^+)$ & $\Xi_{cc}\Xi_{cc}^\ast$ & $-0.0949$ & - \\ \midrule[0.05mm]
        $0(2^+)$ & $\Xi_{cc}\Xi_{cc}$ & $66.5 - i\frac{2.43}{2}$ & $7.01 - i\frac{0.0113}{2}$ \\\midrule[0.05mm]
        \multirow{2}{*}{$1(2^+)$} & \multirow{2}{*}{$\Xi_{cc}\Xi_{cc}$} & $98.0 - i \frac{0.132}{2}$ & \multirow{2}{*}{-} \\
        && $211 - i \frac{2.12}{2}$ & \\
        \bottomrule[0.3mm]
    \end{tabular}\label{tab;mass_XiXi}
\end{table}
\renewcommand{\arraystretch}{1.00}

In this subsection, we show the bound and resonant states of $\Xis\Xis$ with some quantum numbers. 
The channels of $\Xis\Xis$ are summarized in Table.~\ref{tab;Channel_XiXi}. 
The numerical results for $\Xis\Xis$ are summarized in Fig.~\ref{fig;mass_XiXi} and Table~\ref{tab;mass_XiXi}. In the similar way with the case of $\bar{D}^{(\ast)}\Xi_{cc}^{(\ast)}$, we obtain $S$-wave (quasi) bound states.

\begin{figure*}[tb]
    \begin{tabular}{cc}
        $g_\sigma = g^L_\sigma$ & $g_\sigma = g^S_\sigma$ \\
        \includegraphics[width=0.45\linewidth]{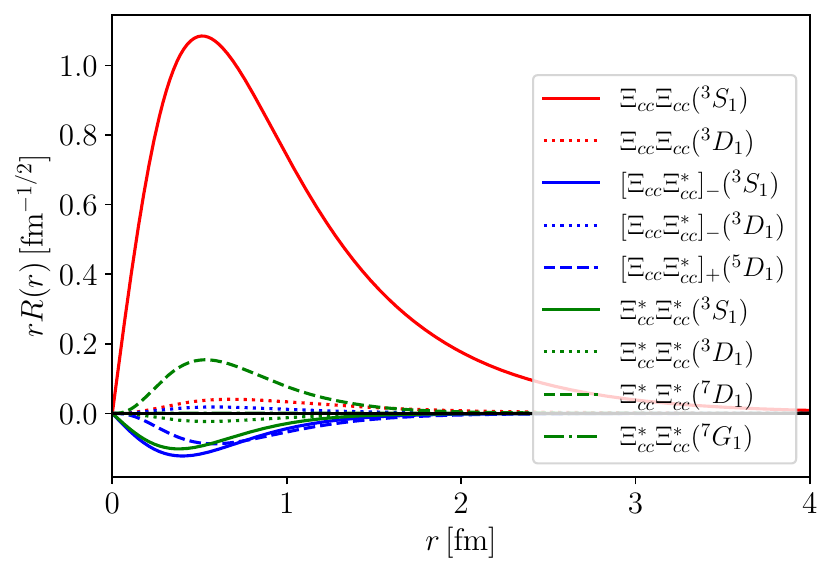}&
        \includegraphics[width=0.45\linewidth]{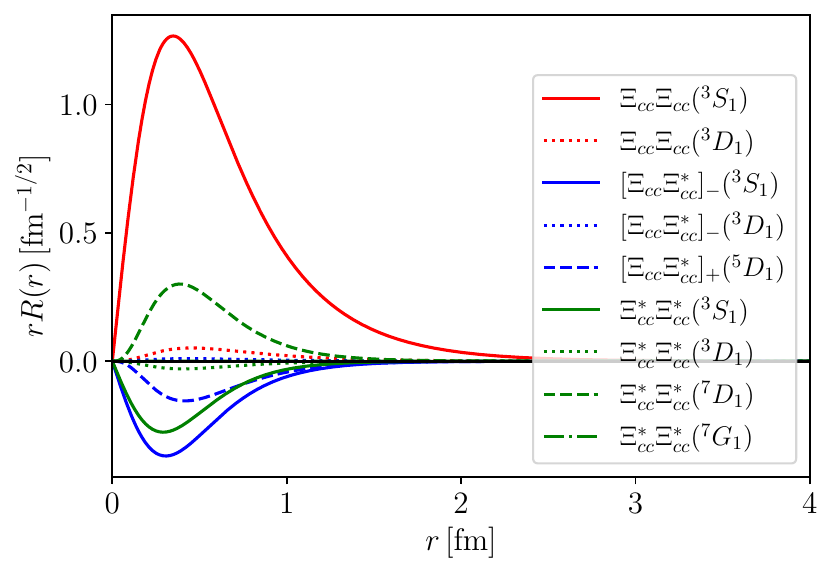}
    \end{tabular}
    \caption{Wavefunctions of $\Xis\Xis$ with $I(J^P) = 0(1^+)$. The left panel shows the wavefunction of $\Xis\Xis$ with $g_\sigma^L$, while right one shows the wavefunction of $\Xis\Xis$ with $g_\sigma^S$. The red, blue and green lines show the wavefunctions for $\Xi_{cc}\Xi_{cc}$, $[\Xi_{cc}\Xi_{cc}^\ast]_\pm$ and $\Xi_{cc}^\ast\Xi^\ast_{cc}$, respectively. }
    \label{fig;wavefunction_XiXi}
\end{figure*}
\renewcommand{\arraystretch}{1.25}
\begin{table}[tb]
    \centering
    \caption{The binding energy $B$, mixing ratios and root mean squared (r.m.s) distances of $\Xis\Xis$ with $0(1^+)$ for each case of $g_\sigma = g^L_\sigma$ and $g_\sigma = g^S_\sigma$. 
    The same conventions as Table~\ref{tab;Properties_DXi} are used. }
    \begin{tabular}{ccc}
        \toprule[0.3mm]
        \multirow{2}{*}{Properties}& \multicolumn{2}{c}{$g_\sigma$}\\
        & $g^L_\sigma = 3.4$ & $g^S_\sigma = 0.76$\\
        \midrule[0.1mm]
        $B$ & $24.9\,\mathrm{MeV}$ & $67.6\,\mathrm{MeV}$ \\
        $\Xi_{cc}\Xi_{cc}(^3S)$ & $96.6\,\%$ & $85.9\,\%$\\
        $\Xi_{cc}\Xi_{cc}(^3D)$ & $0.133\,\%$ & $0.143\,\%$\\
        $[\Xi_{cc}\Xi_{cc}^\ast]_-(^3S)$ & $0.844\,\%$ & $5.94\,\%$\\
        $[\Xi_{cc}\Xi_{cc}^\ast]_-(^3 D)$ & $0.0219\,\%$ & $0\,\%$\\
        $[\Xi_{cc}\Xi_{cc}^\ast]_+(^5D)$ & $0.477\,\%$ & $1.10\,\%$\\       
        $\Xi_{cc}^\ast \Xi_{cc}^\ast(^3S)$ & $0.544\,\%$ & $3.09\,\%$\\ 
        $\Xi_{cc}^\ast \Xi_{cc}^\ast(^3D)$ & $0.0314\,\%$ & $0.0402\,\%$\\ 
        $\Xi_{cc}^\ast \Xi_{cc}^\ast(^7D)$ & $1.34\,\%$ & $3.80\,\%$\\ 
        $\Xi_{cc}^\ast \Xi_{cc}^\ast(^7G)$ & $0\,\%$ & $0\,\%$\\ 
        r.m.s distance & $0.825$ fm  & $0.530$ fm\\
        \bottomrule[0.3mm]
    \end{tabular}\label{tab;Properties_XiXi}
\end{table}
\renewcommand{\arraystretch}{1.00}
\begin{figure}[tbp]
    \centering
    \begin{center}
        \includegraphics[width=0.8\linewidth]{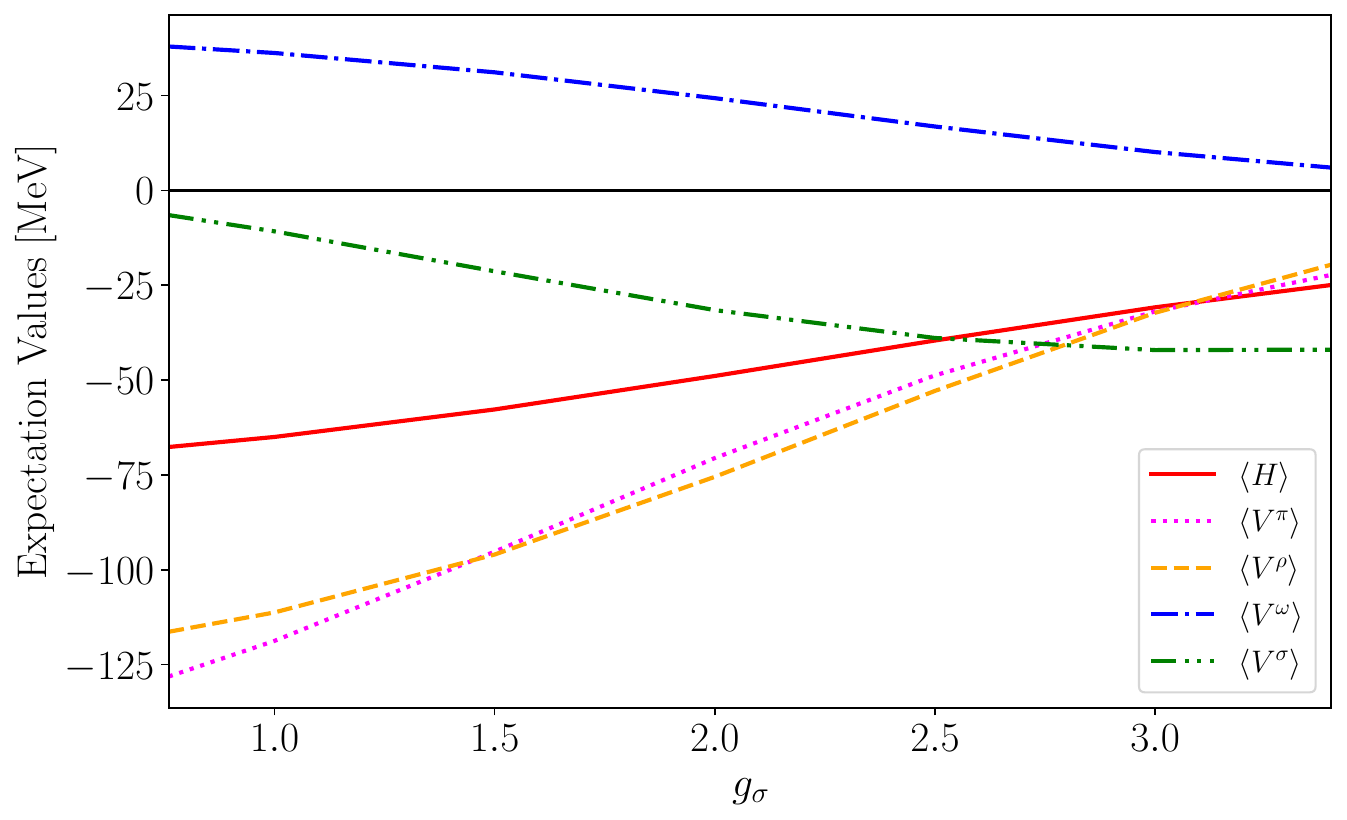}
    \end{center}
    \caption{$g_\sigma$ dependence of the expectation values of the Hamiltonian and OBEP for $\Xis\Xis$ with $0(1^+)$. The same notation as Fig.~\ref{fig;DXi_Expectation_gsigma} is used. }
    \label{fig;XiXi_Expectation_gsigma}
\end{figure}

First, we discuss the bound states of $\Xis\Xis$ with $I(J^P) = 0(1^+)$. 
The wavefunctions of $\Xis\Xis$ with $0(1^+)$ are shown in Fig.~\ref{fig;wavefunction_XiXi}, and their binding energies, mixing ratios and r.m.s distances are summarized in Table.~\ref{tab;Properties_XiXi}. 
{As seen in $\barDs\Xis$} with $0(\frac{1}{2}^-)$, the binding energy of $\Xis\Xis$ with $0(1^+)$ for $g_\sigma = g_\sigma^S$ is larger than for $g_\sigma = g_\sigma^L$. In both cases, the mixing ratio of $\Xis\Xis(^3S)$ is dominant, while that of the other channels differ between the cases of each $g_\sigma$. 
In order to clarify this dependence, we calculate the expectation values of the Hamiltonian and one boson exchange potentials, as shown in Fig.~\ref{fig;XiXi_Expectation_gsigma}. 
As in the case of the bound state of $\barDs\Xis$ with $0(\frac{1}{2}^-)$, the one $\sigma$ exchange potential provides the strongest attraction for $g_\sigma = g_\sigma^L$, whereas for $g_\sigma = g_\sigma^S$ the dominant attraction arises from the one $\pi$ and one $\rho$ exchange potentials.
In particular, the matrix elements $\braket{\Xi_{cc}\Xi_{cc}(^3S)|V^\pi|\Xi_{cc}^\ast \Xi_{cc}^\ast (^7D)}$ and $\braket{\Xi_{cc}\Xi_{cc}(^3S)|V^\rho|\Xi_{cc}\Xi_{cc}(^3S)}$ play important roles in binding $\Xis\Xis(0(1^+))$ with $g_\sigma = g_\sigma^S$. 
Moreover, the strength of $\braket{\Xi_{cc}\Xi_{cc}(^3S)|V^\pi|\Xi_{cc}^\ast \Xi_{cc}^\ast (^7D)}$ indicates that the $\Xi_{cc}^\ast\Xi_{cc}^\ast(^7D_1)$ component provides the third-largest contribution to the channel mixing.

\begin{figure*}[tbp]
    \begin{tabular}{cc}
        $g_\sigma = g^L_\sigma$ & $g_\sigma = g^S_\sigma$ \\
        \includegraphics[width=0.5\linewidth]{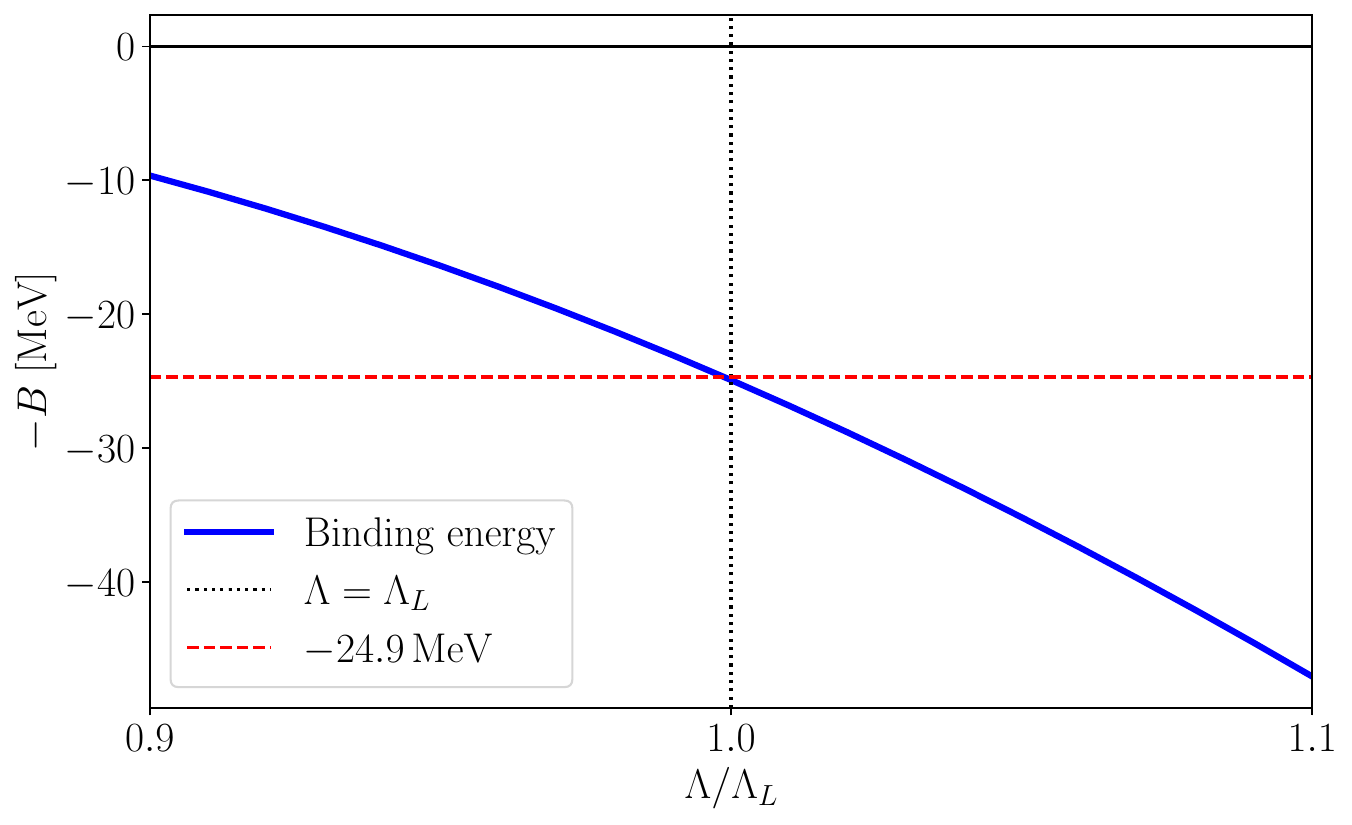}&
        \includegraphics[width=0.5\linewidth]{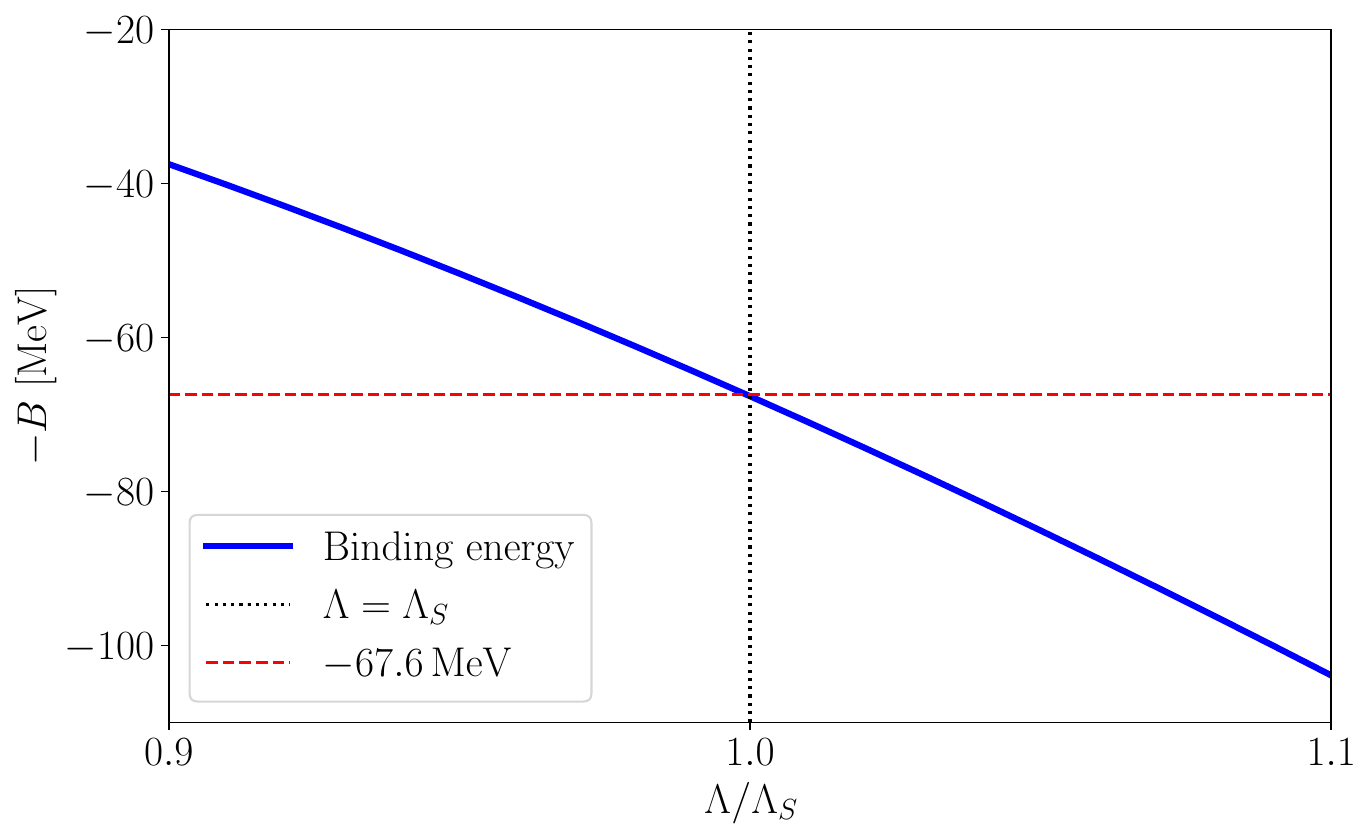}
    \end{tabular}
    \caption{Cutoff dependence of the binding energy of $\Xis\Xis$. The left panel shows the results for $g_\sigma^L$, and the right panel for $g_\sigma^S$. Here, $B$ denotes the binding energy of $\Xis\Xis$.}
    \label{fig;Lambda_dep_XiXi}
\end{figure*}

As the case of $\barDs\Xis$, we investigate the cutoff dependence of binding energies of $\Xis\Xis$ with $g^L_\sigma$ and $g^S_\sigma$ in Fig.~\ref{fig;Lambda_dep_XiXi}. 
This result shows that the larger the cutoff parameter becomes, the larger the binding energy of $\Xis\Xis$ with $g_\sigma^L$ and $g_\sigma^S$ becomes.

We also discuss the bound states of $\Xis\Xis$ with $I = 1$. In this case, bound states of $\Xis\Xis$ with $1(0^+)$ and $1(1^+)$ are found, and their properties are discussed below. 
The mixing ratios of $\Xi_{cc}\Xi_{cc}(^1S_0)$ for $\Xis\Xis(1(0^+))$ and $[\Xi_{cc}\Xi^\ast_{cc}]_+(^3S_1)$ for $\Xis\Xis(1(1^+))$ are about $99\,\%$. This indicates that the one $\sigma$ exchange potential provides the dominant attraction that is responsible for binding these states. 
On the other hand, for $g_\sigma = g_\sigma^S$, bound states of $\Xis\Xis$ with $I=1$ are not obtained. This is because the one $\sigma$ exchange potential becomes weaker for $g_\sigma = g_\sigma^S$ than for $g_\sigma = g_\sigma^L$, and the one $\pi$ and one $\rho$ exchange potentials cannot generate sufficient attractive force to bind $\Xis\Xis$ with $I=1$.

Next, we discuss the resonances of $\Xis\Xis$. All resonances are identified as Feshbach resonances because bound states of the upper channels appear when the lower channels are cut. 
Moreover, most quasi bound states of $\Xi_{cc}\Xi^\ast_{cc}$ obtained with $g_\sigma^S$ are bound more deeply than those obtained with $g_\sigma^L$ for the same reasons as in the case of bound states of $\Xis\Xis$. 
However, no quasi bound state of $\Xi^\ast_{cc}\Xi^\ast_{cc}$ with $I(J^P) = 0(1^+)$ is found for $g_\sigma = g_\sigma^S$. This is because the one $\sigma$ exchange potential is weaker for $g_\sigma = g_\sigma^S$ than for $g_\sigma = g_\sigma^L$, and the one $\pi$ and one $\rho$ exchange potentials do not provide enough attraction to form a bound state. 

Let us comment on the $g_\sigma$ dependence of the results. 
Throughout this work, we have examined the dependence of the results on $g_\sigma$.
Although all parameter sets reproduce the binding energy of $T_{cc}$, it is nevertheless intriguing that the resulting predictions differ so significantly.
In the future, if $\barDs\Xis$ and $\Xis\Xis$ are observed experimentally or studied in lattice QCD, they may provide useful constraints for determining the model parameters.

\section{Summary}\label{sec;summary}

In this paper, we have investigated $\barDs\Xis$ and $\Xis\Xis$ molecules, respecting the superflavor and heavy quark spin symmetries. 
These states are superflavor partners of a $D^{(\ast)}D^{(\ast)}$ molecule which is a candidate of the structure of $T_{cc}$, which has been reported by the LHCb in Refs.~\cite{LHCb:2021vvq,LHCb:2021chn}. 
Thus, in our previous studies~\cite{Sakai:2023syt,Sakai:2025djx}, we studied $T_{cc}$ as a $D^{(\ast)}D^{(\ast)}$ molecule based on the one boson exchange potential, respecting the heavy quark symmetry. 
With the help of the superflavor symmetry, we have applied our models for $D^{(\ast)}D^{(\ast)}$ to the $\barDs\Xis$ and $\Xis\Xis$ systems, using the same parameters with $T_{cc}$. Most parameters are determined by the experimental data or Lattice QCD, while the $g_\sigma$ coupling constant is uncertain. However, we consider two typical values of $g_\sigma$: $g_\sigma^L = 3.4$ and $g_\sigma^S = 0.76$. 
Then, we investigate the $g_\sigma$ dependence of $\barDs\Xis$ and $\Xis\Xis$ states.

First, the cutoff parameter $\Lambda$ has been determined to reproduce the binding energy of $T_{cc}$ for each $g_\sigma$. 
It was found that the cutoff for $g_\sigma = g_\sigma^S$ is larger than that for $g_\sigma = g_\sigma^L$. This is because for $g_\sigma = g_\sigma^L$ the one $\sigma$ exchange potential generates enough attractive force to bind $T_{cc}$, while for $g_\sigma = g^S_\sigma$ the contribution of one $\sigma$ exchange force becomes weaker, then one $\pi$ and one $\rho$ exchange potentials should earn enough attractive force. 

We have obtained many bound and resonant states of $\barDs\Xis$ with some quantum numbers. 
The bound states of $\barDs\Xis$ were obtained only for $I(J^P)=0(\frac{1}{2}^-)$. 
This is because only the $\barDs\Xis$ state with $J^P = \frac{1}{2}^-$ has $S$-wave $\bar{D}\Xi_{cc}$ channel. 
The binding energy of $\barDs\Xis$ for $g_\sigma = g_\sigma^S$ is smaller than that for $g_\sigma = g_\sigma^L$ because the one $\pi$ and $\rho$ exchange potentials provide stronger attractive force. 
For both $g_\sigma$ cases, the channel of $\bar{D}\Xi_{cc}(^2S)$ is dominant. However, the {mixing ratios} of the other channels depends on the value of $g_\sigma$.  For $g_\sigma = g_\sigma^L$ the contributions of the other channels are suppressed, while for $g_\sigma = g_\sigma^S$ some channels are enhanced by the off-diagonal components of one $\pi$ and one $\rho$ exchange potentials. 
As for resonant states, many resonances with not only $J^P = \frac{1}{2}^-$ but also $J^P = \frac{3}{2}^-, \frac{5}{2}^-$ were obtained because a large number of $\bar{D}^\ast \Xi^\ast_{cc}$ channels contribute to the higher-spin states. 
All resonances are found to be Feshbach resonances. In the isoscalar states, most resonances for $g_\sigma = g_\sigma^S$ are more deeply bound than those for $g_\sigma = g_\sigma^L$. For the $\bar{D}^*\Xi_{cc}^*$ quasi bound state with $I(J^P)=0(\frac{3}{2}^-)$, however, we obtain the state only for $g_\sigma=g_\sigma^L$, while no bound state is found for $g_\sigma=g_\sigma^S$. This is because the one $\pi$ and one $\rho$ exchange potentials cannot generate enough attractive force. 
For the isovector state, we obtained the resonance only for $I(J^P)=1(\frac{1}{2}^-)$ with $g_\sigma = g_\sigma^L$, where the one $\sigma$ exchange potential plays an important role to produce an attraction.

We also have investigated bound and resonant states of $\Xis\Xis$ with some quantum numbers. As well as the $\barDs\Xis$ states, many bound and resonant states of $\Xis\Xis$ were obtained. The $\Xis\Xis$ states with $I=0$ were obtained for each $g_\sigma$, while the $\Xis\Xis$ states with $I=1$ were obtained only for $g_\sigma = g_\sigma^L$. 
As for the bound states of $\Xis\Xis$ with $I(J^P) = 0(1^+)$, the binding energy for $g_\sigma = g_\sigma^S$ is larger than that for $g_\sigma = g_\sigma^L$ because the one $\pi$ and one $\rho$ exchange potentials earn stronger attractive forces. Moreover, all resonances of $\Xis\Xis$ were found to be Feshbach resonances. As for the $\Xis\Xis$ resonances with $I=0$, quasi bound states of $\Xi_{cc}\Xi^\ast_{cc}$ for $g_\sigma = g_\sigma^S$ are more bound than those for $g_\sigma^L$. 
However, the quasi bound state of $\Xi^\ast_{cc}\Xi^\ast_{cc}$ were obtained only for $g_\sigma = g_\sigma^L$. 

{Once experimental data and Lattice data for $\barDs\Xis$ and $\Xis\Xis$ become available, they could provide significant constraints on the coupling constants.
}

\section*{Acknowledgement}
This work is supported by the Grant-in-Aid for JSPS Fellows No. 25KJ1433 (M.S.) and RCNP Collaboration Research Network program as the project number COREnet-056 (Y.Y.).

\bibliography{apssamp}

@article{Casalbuoni:1996pg,
    author = "Casalbuoni, R. and Deandrea, A. and Di Bartolomeo, N. and Gatto, Raoul and Feruglio, F. and Nardulli, G.",
    title = "{Phenomenology of heavy meson chiral Lagrangians}",
    eprint = "hep-ph/9605342",
    archivePrefix = "arXiv",
    reportNumber = "UGVA-DPT-1996-05-928, BARI-TH-96-237",
    doi = "10.1016/S0370-1573(96)00027-0",
    journal = "Phys. Rept.",
    volume = "281",
    pages = "145--238",
    year = "1997"
}

@article{Ohkoda:2012hv,
    author = "Ohkoda, S. and Yamaguchi, Y. and Yasui, S. and Sudoh, K. and Hosaka, A.",
    title = "{Exotic mesons with double charm and bottom flavor}",
    eprint = "1202.0760",
    archivePrefix = "arXiv",
    primaryClass = "hep-ph",
    doi = "10.1103/PhysRevD.86.034019",
    journal = "Phys. Rev. D",
    volume = "86",
    pages = "034019",
    year = "2012"
}

@article{LHCb:2021vvq,
    author = "Aaij, Roel and others",
    collaboration = "LHCb",
    title = "{Observation of an exotic narrow doubly charmed tetraquark}",
    eprint = "2109.01038",
    archivePrefix = "arXiv",
    primaryClass = "hep-ex",
    reportNumber = "CERN-EP-2021-165, LHCb-PAPER-2021-031",
    doi = "10.1038/s41567-022-01614-y",
    journal = "Nature Phys.",
    volume = "18",
    number = "7",
    pages = "751--754",
    year = "2022"
}

@article{Isola:2003fh,
    author = "Isola, Claudia and Ladisa, Massimo and Nardulli, Giuseppe and Santorelli, Pietro",
    title = "{Charming penguins in B ---\ensuremath{>} K* pi, K(rho, omega, phi) decays}",
    eprint = "hep-ph/0307367",
    archivePrefix = "arXiv",
    reportNumber = "BARI-TH-469-03, DSF-2003-26, DSF-2003-26-(NAPOLI)",
    doi = "10.1103/PhysRevD.68.114001",
    journal = "Phys. Rev. D",
    volume = "68",
    pages = "114001",
    year = "2003"
}

@article{CLEO:2001foe,
    author = "Ahmed, S. and others",
    collaboration = "CLEO",
    title = "{First measurement of Gamma(D*+)}",
    eprint = "hep-ex/0108013",
    archivePrefix = "arXiv",
    reportNumber = "CLNS-01-1740",
    doi = "10.1103/PhysRevLett.87.251801",
    journal = "Phys. Rev. Lett.",
    volume = "87",
    pages = "251801",
    year = "2001"
}

@article{Belle:2003nnu,
    author = "Choi, S. K. and others",
    collaboration = "Belle",
    title = "{Observation of a narrow charmonium-like state in exclusive $B^\pm \to K^\pm \pi^+ \pi^- J/\psi$ decays}",
    eprint = "hep-ex/0309032",
    archivePrefix = "arXiv",
    doi = "10.1103/PhysRevLett.91.262001",
    journal = "Phys. Rev. Lett.",
    volume = "91",
    pages = "262001",
    year = "2003"
}

@article{LHCb:2015yax,
    author = "Aaij, Roel and others",
    collaboration = "LHCb",
    title = "{Observation of $J/\psi p$ Resonances Consistent with Pentaquark States in $\Lambda_b^0 \to J/\psi K^- p$ Decays}",
    eprint = "1507.03414",
    archivePrefix = "arXiv",
    primaryClass = "hep-ex",
    reportNumber = "CERN-PH-EP-2015-153, LHCB-PAPER-2015-029",
    doi = "10.1103/PhysRevLett.115.072001",
    journal = "Phys. Rev. Lett.",
    volume = "115",
    pages = "072001",
    year = "2015"
}

@article{LHCb:2019kea,
    author = "Aaij, Roel and others",
    collaboration = "LHCb",
    title = "{Observation of a narrow pentaquark state, $P_c(4312)^+$, and of two-peak structure of the $P_c(4450)^+$}",
    eprint = "1904.03947",
    archivePrefix = "arXiv",
    primaryClass = "hep-ex",
    reportNumber = "LHCb-PAPER-2019-014 CERN-EP-2019-058",
    doi = "10.1103/PhysRevLett.122.222001",
    journal = "Phys. Rev. Lett.",
    volume = "122",
    number = "22",
    pages = "222001",
    year = "2019"
}

@article{LHCb:2020bwg,
    author = "Aaij, Roel and others",
    collaboration = "LHCb",
    title = "{Observation of structure in the $J /\psi$ -pair mass spectrum}",
    eprint = "2006.16957",
    archivePrefix = "arXiv",
    primaryClass = "hep-ex",
    reportNumber = "CERN-EP-2020-115, LHCb-PAPER-2020-011",
    doi = "10.1016/j.scib.2020.08.032",
    journal = "Sci. Bull.",
    volume = "65",
    number = "23",
    pages = "1983--1993",
    year = "2020"
}

@inproceedings{Grinstein:1995uv,
    author = "Grinstein, Benjamin",
    title = "{An Introduction to heavy mesons}",
    booktitle = "{6th Mexican School of Particles and Fields}",
    eprint = "hep-ph/9508227",
    archivePrefix = "arXiv",
    reportNumber = "UCSD-PTH-95-05",
    pages = "122--184",
    month = "8",
    year = "1995"
}

@article{Neubert:1993mb,
    author = "Neubert, Matthias",
    title = "{Heavy quark symmetry}",
    eprint = "hep-ph/9306320",
    archivePrefix = "arXiv",
    reportNumber = "SLAC-PUB-6263",
    doi = "10.1016/0370-1573(94)90091-4",
    journal = "Phys. Rept.",
    volume = "245",
    pages = "259--396",
    year = "1994"
}

@article{Li:2012ss,
    author = "Li, Ning and Sun, Zhi-Feng and Liu, Xiang and Zhu, Shi-Lin",
    title = "{Coupled-channel analysis of the possible $D^{(*)}D^{(*)}, \overline{B}^{(*)}\overline{B}^{(*)}$ and $D^{(*)}\overline{B}^{(*)}$ molecular states}",
    eprint = "1211.5007",
    archivePrefix = "arXiv",
    primaryClass = "hep-ph",
    doi = "10.1103/PhysRevD.88.114008",
    journal = "Phys. Rev. D",
    volume = "88",
    number = "11",
    pages = "114008",
    year = "2013"
}

@article{Wang:2021ajy,
    author = "Wang, Fu-Lai and Chen, Rui and Liu, Xiang",
    title = "{A new group of doubly charmed molecule with T-doublet charmed meson pair}",
    eprint = "2111.00208",
    archivePrefix = "arXiv",
    primaryClass = "hep-ph",
    doi = "10.1016/j.physletb.2022.137502",
    journal = "Phys. Lett. B",
    volume = "835",
    pages = "137502",
    year = "2022"
}

@article{Wang:2021yld,
    author = "Wang, Fu-Lai and Liu, Xiang",
    title = "{Investigating new type of doubly charmed molecular tetraquarks composed of charmed mesons in the H and T doublets}",
    eprint = "2108.09925",
    archivePrefix = "arXiv",
    primaryClass = "hep-ph",
    doi = "10.1103/PhysRevD.104.094030",
    journal = "Phys. Rev. D",
    volume = "104",
    number = "9",
    pages = "094030",
    year = "2021"
}

@article{Ren:2021dsi,
    author = "Ren, Huimin and Wu, Fan and Zhu, Ruilin",
    title = "{Hadronic Molecule Interpretation of Tcc+ and Its Beauty Partners}",
    eprint = "2109.02531",
    archivePrefix = "arXiv",
    primaryClass = "hep-ph",
    doi = "10.1155/2022/9103031",
    journal = "Adv. High Energy Phys.",
    volume = "2022",
    pages = "9103031",
    year = "2022"
}

@article{Lyu:2023xro,
    author = "Lyu, Yan and Aoki, Sinya and Doi, Takumi and Hatsuda, Tetsuo and Ikeda, Yoichi and Meng, Jie",
    title = "{Doubly Charmed Tetraquark Tcc+ from Lattice QCD near Physical Point}",
    eprint = "2302.04505",
    archivePrefix = "arXiv",
    primaryClass = "hep-lat",
    reportNumber = "RIKEN-iTHEMS-Report-23, YITP-23-14",
    doi = "10.1103/PhysRevLett.131.161901",
    journal = "Phys. Rev. Lett.",
    volume = "131",
    number = "16",
    pages = "161901",
    year = "2023"
}

@article{Ikeda:2013vwa,
    author = "Ikeda, Yoichi and Charron, Bruno and Aoki, Sinya and Doi, Takumi and Hatsuda, Tetsuo and Inoue, Takashi and Ishii, Noriyoshi and Murano, Keiko and Nemura, Hidekatsu and Sasaki, Kenji",
    title = "{Charmed tetraquarks $T_{cc}$ and $T_{cs}$ from dynamical lattice QCD simulations}",
    eprint = "1311.6214",
    archivePrefix = "arXiv",
    primaryClass = "hep-lat",
    reportNumber = "RIKEN-QHP-105, YITP-13-119",
    doi = "10.1016/j.physletb.2014.01.002",
    journal = "Phys. Lett. B",
    volume = "729",
    pages = "85--90",
    year = "2014"
}

@article{Chen:2022asf,
    author = "Chen, Hua-Xing and Chen, Wei and Liu, Xiang and Liu, Yan-Rui and Zhu, Shi-Lin",
    title = "{An updated review of the new hadron states}",
    eprint = "2204.02649",
    archivePrefix = "arXiv",
    primaryClass = "hep-ph",
    doi = "10.1088/1361-6633/aca3b6",
    journal = "Rept. Prog. Phys.",
    volume = "86",
    number = "2",
    pages = "026201",
    year = "2023"
}

@article{LHCb:2021auc,
    author = "Aaij, Roel and others",
    collaboration = "LHCb",
    title = "{Study of the doubly charmed tetraquark $T_{cc}^{+}$}",
    eprint = "2109.01056",
    archivePrefix = "arXiv",
    primaryClass = "hep-ex",
    reportNumber = "CERN-EP-2021-169, LHCb-PAPER-2021-032",
    doi = "10.1038/s41467-022-30206-w",
    journal = "Nature Commun.",
    volume = "13",
    number = "1",
    pages = "3351",
    year = "2022"
}

@article{Padmanath:2022cvl,
    author = "Padmanath, M. and Prelovsek, S.",
    title = "{Signature of a Doubly Charm Tetraquark Pole in DD* Scattering on the Lattice}",
    eprint = "2202.10110",
    archivePrefix = "arXiv",
    primaryClass = "hep-lat",
    reportNumber = "MITP/22-018",
    doi = "10.1103/PhysRevLett.129.032002",
    journal = "Phys. Rev. Lett.",
    volume = "129",
    number = "3",
    pages = "032002",
    year = "2022"
}

@article{Tornqvist:1993ng,
    author = "Tornqvist, Nils A.",
    title = "{From the deuteron to deusons, an analysis of deuteron - like meson meson bound states}",
    eprint = "hep-ph/9310247",
    archivePrefix = "arXiv",
    reportNumber = "HU-SEFT-R-1993-12",
    doi = "10.1007/BF01413192",
    journal = "Z. Phys. C",
    volume = "61",
    pages = "525--537",
    year = "1994"
}

@article{Sakai:2023syt,
    author = "Sakai, Manato and Yamaguchi, Yasuhiro",
    title = "{Analysis of Tcc and Tbb based on the hadronic molecular model and their spin multiplets}",
    eprint = "2312.08663",
    archivePrefix = "arXiv",
    primaryClass = "hep-ph",
    doi = "10.1103/PhysRevD.109.054016",
    journal = "Phys. Rev. D",
    volume = "109",
    number = "5",
    pages = "054016",
    year = "2024"
}

@article{Myo:2014ypa,
    author = "Myo, Takayuki and Kikuchi, Yuma and Masui, Hiroshi and Kat\={o}, Kiyoshi",
    title = "{Recent development of complex scaling method for many-body resonances and continua in light nuclei}",
    eprint = "1410.4356",
    archivePrefix = "arXiv",
    primaryClass = "nucl-th",
    doi = "10.1016/j.ppnp.2014.08.001",
    journal = "Prog. Part. Nucl. Phys.",
    volume = "79",
    pages = "1--56",
    year = "2014"
}

@article{Suzuki:2005wv,
    author = "Suzuki, Ryusuke and Myo, Takayuki and Kato, Kiyoshi",
    editor = "Kalantar-Nayestanaki, Nasser and Timmermans, Rob G. E. and Bakker, Bernard L. G.",
    title = "{Level density in complex scaling method}",
    eprint = "nucl-th/0502012",
    archivePrefix = "arXiv",
    doi = "10.1063/1.1933001",
    journal = "AIP Conf. Proc.",
    volume = "768",
    number = "1",
    pages = "455",
    year = "2005"
}

@article{Myo:2020rni,
    author = "Myo, Takayuki and Kato, Kiyoshi",
    title = "{Complex scaling: Physics of unbound light nuclei and perspective}",
    eprint = "2007.12172",
    archivePrefix = "arXiv",
    primaryClass = "nucl-th",
    doi = "10.1093/ptep/ptaa101",
    journal = "PTEP",
    volume = "2020",
    number = "12",
    pages = "12A101",
    year = "2020"
}

@article{BaBar:2005hhc,
    author = "Aubert, Bernard and others",
    collaboration = "BaBar",
    title = "{Observation of a broad structure in the $\pi^+ \pi^- J/\psi$ mass spectrum around 4.26-GeV/c$^2$}",
    eprint = "hep-ex/0506081",
    archivePrefix = "arXiv",
    reportNumber = "BABAR-PUB-05-29, SLAC-PUB-11320, BABAR-PUB-05-029",
    doi = "10.1103/PhysRevLett.95.142001",
    journal = "Phys. Rev. Lett.",
    volume = "95",
    pages = "142001",
    year = "2005"
}

@article{Belle:2007dxy,
    author = "Yuan, C. Z. and others",
    collaboration = "Belle",
    title = "{Measurement of e+ e- ---\ensuremath{>} pi+ pi- J/psi cross-section via initial state radiation at Belle}",
    eprint = "0707.2541",
    archivePrefix = "arXiv",
    primaryClass = "hep-ex",
    reportNumber = "BELLE-PREPRINT-2007-31, KEK-PREPRINT-2007-23",
    doi = "10.1103/PhysRevLett.99.182004",
    journal = "Phys. Rev. Lett.",
    volume = "99",
    pages = "182004",
    year = "2007"
}

@article{Belle:2007hrb,
    author = "Choi, S. K. and others",
    editor = "Son, Dongchul and Oh, Sun Kun",
    collaboration = "Belle",
    title = "{Observation of a resonance-like structure in the $pi^\pm \psi^\prime$ mass distribution in exclusive $B \to K \pi^\pm \psi^\prime$ decays}",
    eprint = "0708.1790",
    archivePrefix = "arXiv",
    primaryClass = "hep-ex",
    reportNumber = "BELLE-CONF-0773",
    doi = "10.1103/PhysRevLett.100.142001",
    journal = "Phys. Rev. Lett.",
    volume = "100",
    pages = "142001",
    year = "2008"
}

@article{BaBar:2008bxw,
    author = "Aubert, Bernard and others",
    collaboration = "BaBar",
    title = "{Search for the Z(4430)- at BABAR}",
    eprint = "0811.0564",
    archivePrefix = "arXiv",
    primaryClass = "hep-ex",
    reportNumber = "SLAC-PUB-13437, BABAR-PUB-08-045",
    doi = "10.1103/PhysRevD.79.112001",
    journal = "Phys. Rev. D",
    volume = "79",
    pages = "112001",
    year = "2009"
}

@article{BESIII:2020qkh,
    author = "Ablikim, Medina and others",
    collaboration = "BESIII",
    title = "{Observation of a Near-Threshold Structure in the $K^+$ Recoil-Mass Spectra in $e^+e^- \rightarrow K^+(D_s^-D^{*0}+D_s^{*-}D^0$)}",
    eprint = "2011.07855",
    archivePrefix = "arXiv",
    primaryClass = "hep-ex",
    doi = "10.1103/PhysRevLett.126.102001",
    journal = "Phys. Rev. Lett.",
    volume = "126",
    number = "10",
    pages = "102001",
    year = "2021"
}

@article{LHCb:2020jpq,
    author = "Aaij, Roel and others",
    collaboration = "LHCb",
    title = "{Evidence of a $J/\psi\Lambda$ structure and observation of excited $\Xi^-$ states in the $\Xi^-_b \to J/\psi\Lambda K^-$ decay}",
    eprint = "2012.10380",
    archivePrefix = "arXiv",
    primaryClass = "hep-ex",
    reportNumber = "LHCb-PAPER-2020-039, CERN-EP-2020-233",
    doi = "10.1016/j.scib.2021.02.030",
    journal = "Sci. Bull.",
    volume = "66",
    pages = "1278--1287",
    year = "2021"
}

@article{LHCb:2021chn,
    author = "Aaij, Roel and others",
    collaboration = "LHCb",
    title = "{Evidence for a new structure in the $J/\psi p$ and $J/\psi \bar{p}$ systems in $B_s^0 \to J/\psi p \bar{p}$ decays}",
    eprint = "2108.04720",
    archivePrefix = "arXiv",
    primaryClass = "hep-ex",
    reportNumber = "LHCb-PAPER-2021-018, CERN-EP-2021-150",
    doi = "10.1103/PhysRevLett.128.062001",
    journal = "Phys. Rev. Lett.",
    volume = "128",
    number = "6",
    pages = "062001",
    year = "2022"
}

@article{Asanuma:2023atv,
    author = "Asanuma, Tatsuya and Yamaguchi, Yasuhiro and Harada, Masayasu",
    title = "{Analysis of DD* and D\textasciimacron{}(*)\ensuremath{\Xi}cc(*) molecule by one boson exchange model based on heavy quark symmetry}",
    eprint = "2311.04695",
    archivePrefix = "arXiv",
    primaryClass = "hep-ph",
    doi = "10.1103/PhysRevD.110.074030",
    journal = "Phys. Rev. D",
    volume = "110",
    number = "7",
    pages = "074030",
    year = "2024"
}

@article{Liu:2019stu,
    author = "Liu, Ming-Zhu and Wu, Tian-Wei and Pavon Valderrama, Manuel and Xie, Ju-Jun and Geng, Li-Sheng",
    title = "{Heavy-quark spin and flavor symmetry partners of the X(3872) revisited: What can we learn from the one boson exchange model?}",
    eprint = "1902.03044",
    archivePrefix = "arXiv",
    primaryClass = "hep-ph",
    doi = "10.1103/PhysRevD.99.094018",
    journal = "Phys. Rev. D",
    volume = "99",
    number = "9",
    pages = "094018",
    year = "2019"
}

@article{Hiyama:2003cu,
    author = "Hiyama, E. and Kino, Y. and Kamimura, M.",
    title = "{Gaussian expansion method for few-body systems}",
    doi = "10.1016/S0146-6410(03)90015-9",
    journal = "Prog. Part. Nucl. Phys.",
    volume = "51",
    pages = "223--307",
    year = "2003"
}

@article{Hiyama:2018ivm,
    author = "Hiyama, Emiko and Kamimura, Masayasu",
    title = "{Study of various few-body systems using Gaussian expansion method (GEM)}",
    eprint = "1809.02619",
    archivePrefix = "arXiv",
    primaryClass = "nucl-th",
    doi = "10.1007/s11467-018-0828-5",
    journal = "Front. Phys. (Beijing)",
    volume = "13",
    number = "6",
    pages = "132106",
    year = "2018"
}

@article{Grinstein:1992qt,
    author = "Grinstein, Benjamin and Jenkins, Elizabeth Ellen and Manohar, Aneesh V. and Savage, Martin J. and Wise, Mark B.",
    title = "{Chiral perturbation theory for f D(s) / f D and B B(s) / B B}",
    eprint = "hep-ph/9204207",
    archivePrefix = "arXiv",
    reportNumber = "UCSD-PTH-92-05, CALT-68-1768, SSCL-PREPRINT-025",
    doi = "10.1016/0550-3213(92)90248-A",
    journal = "Nucl. Phys. B",
    volume = "380",
    pages = "369--376",
    year = "1992"
}

@article{Hu:2005gf,
    author = "Hu, Jie and Mehen, Thomas",
    title = "{Chiral Lagrangian with heavy quark-diquark symmetry}",
    eprint = "hep-ph/0511321",
    archivePrefix = "arXiv",
    reportNumber = "JLAB-THY-05-452",
    doi = "10.1103/PhysRevD.73.054003",
    journal = "Phys. Rev. D",
    volume = "73",
    pages = "054003",
    year = "2006"
}

@article{Sakai:2025djx,
    author = "Sakai, Manato and Yamaguchi, Yasuhiro",
    title = "{Analysis of bound and resonant states of doubly heavy tetraquarks with spin J{\ensuremath{\leq}}2}",
    eprint = "2503.11134",
    archivePrefix = "arXiv",
    primaryClass = "hep-ph",
    doi = "10.1103/jx4b-f1by",
    journal = "Phys. Rev. D",
    volume = "112",
    number = "3",
    pages = "034038",
    year = "2025"
}

@article{ParticleDataGroup:2024cfk,
    author = "Navas, S. and others",
    collaboration = "Particle Data Group",
    title = "{Review of particle physics}",
    doi = "10.1103/PhysRevD.110.030001",
    journal = "Phys. Rev. D",
    volume = "110",
    number = "3",
    pages = "030001",
    year = "2024"
}

@article{Georgi:1990ak,
    author = "Georgi, Howard and Wise, Mark B.",
    title = "{Superflavor Symmetry for Heavy Particles}",
    reportNumber = "HUTP-90-A022, CALT-68-1629",
    doi = "10.1016/0370-2693(90)90851-V",
    journal = "Phys. Lett. B",
    volume = "243",
    pages = "279--283",
    year = "1990"
}

@article{Savage:1990di,
    author = "Savage, Martin J. and Wise, Mark B.",
    title = "{Spectrum of baryons with two heavy quarks}",
    reportNumber = "CALT-68-1652",
    doi = "10.1016/0370-2693(90)90035-5",
    journal = "Phys. Lett. B",
    volume = "248",
    pages = "177--180",
    year = "1990"
}

@article{Fleming:2005pd,
    author = "Fleming, Sean and Mehen, Thomas",
    title = "{Doubly heavy baryons, heavy quark-diquark symmetry and NRQCD}",
    eprint = "hep-ph/0509313",
    archivePrefix = "arXiv",
    reportNumber = "JLAB-THY-05-415",
    doi = "10.1103/PhysRevD.73.034502",
    journal = "Phys. Rev. D",
    volume = "73",
    pages = "034502",
    year = "2006"
}

@article{Nagatsuka:2025szy,
    author = "Nagatsuka, Masato and Sasaki, Shoichi",
    title = "{Lattice study of scattering phase shifts for DD* and BB* systems using twisted boundary conditions: Search for bound state formation}",
    eprint = "2507.20712",
    archivePrefix = "arXiv",
    primaryClass = "hep-lat",
    doi = "10.1103/ksmm-8d7r",
    journal = "Phys. Rev. D",
    volume = "112",
    number = "11",
    pages = "114510",
    year = "2025"
}

@article{Inoue:2010es,
    author = "Inoue, Takashi and Ishii, Noriyoshi and Aoki, Sinya and Doi, Takumi and Hatsuda, Tetsuo and Ikeda, Yoichi and Murano, Keiko and Nemura, Hidekatsu and Sasaki, Kenji",
    collaboration = "HAL QCD",
    title = "{Bound H-dibaryon in Flavor SU(3) Limit of Lattice QCD}",
    eprint = "1012.5928",
    archivePrefix = "arXiv",
    primaryClass = "hep-lat",
    doi = "10.1103/PhysRevLett.106.162002",
    journal = "Phys. Rev. Lett.",
    volume = "106",
    pages = "162002",
    year = "2011"
}

@article{Ishii:2012ssm,
    author = "Ishii, Noriyoshi and Aoki, Sinya and Doi, Takumi and Hatsuda, Tetsuo and Ikeda, Yoichi and Inoue, Takashi and Murano, Keiko and Nemura, Hidekatsu and Sasaki, Kenji",
    collaboration = "HAL QCD",
    title = "{Hadron{\textendash}hadron interactions from imaginary-time Nambu{\textendash}Bethe{\textendash}Salpeter wave function on the lattice}",
    eprint = "1203.3642",
    archivePrefix = "arXiv",
    primaryClass = "hep-lat",
    doi = "10.1016/j.physletb.2012.04.076",
    journal = "Phys. Lett. B",
    volume = "712",
    pages = "437--441",
    year = "2012"
}

@article{Aoki:2009ji,
    author = "Aoki, Sinya and Hatsuda, Tetsuo and Ishii, Noriyoshi",
    title = "{Theoretical Foundation of the Nuclear Force in QCD and its applications to Central and Tensor Forces in Quenched Lattice QCD Simulations}",
    eprint = "0909.5585",
    archivePrefix = "arXiv",
    primaryClass = "hep-lat",
    reportNumber = "TKYNT-09-19, UTHEP-591",
    doi = "10.1143/PTP.123.89",
    journal = "Prog. Theor. Phys.",
    volume = "123",
    pages = "89--128",
    year = "2010"
}

@article{Machleidt:1987hj,
    author = "Machleidt, R. and Holinde, K. and Elster, C.",
    title = "{The Bonn Meson Exchange Model for the Nucleon Nucleon Interaction}",
    doi = "10.1016/S0370-1573(87)80002-9",
    journal = "Phys. Rept.",
    volume = "149",
    pages = "1--89",
    year = "1987"
}

@article{Machleidt:2000ge,
    author = "Machleidt, R.",
    title = "{The High precision, charge dependent Bonn nucleon-nucleon potential (CD-Bonn)}",
    eprint = "nucl-th/0006014",
    archivePrefix = "arXiv",
    doi = "10.1103/PhysRevC.63.024001",
    journal = "Phys. Rev. C",
    volume = "63",
    pages = "024001",
    year = "2001"
}

@article{Sekihara:2023ihc,
    author = "Sekihara, Takayasu and Hashiguchi, Taishi",
    title = "{Reexamination of the short-range baryon-baryon potentials in the constituent quark model}",
    eprint = "2304.13877",
    archivePrefix = "arXiv",
    primaryClass = "nucl-th",
    doi = "10.1103/PhysRevC.108.065202",
    journal = "Phys. Rev. C",
    volume = "108",
    number = "6",
    pages = "065202",
    year = "2023"
}

@article{Oka:2000wj,
    author = "Oka, Makoto and Shimizu, K. and Yazaki, K.",
    title = "{Quark cluster model of baryon baryon interaction}",
    doi = "10.1143/PTPS.137.1",
    journal = "Prog. Theor. Phys. Suppl.",
    volume = "137",
    pages = "1--20",
    year = "2000"
}

@article{Oka:1981ri,
    author = "Oka, M. and Yazaki, K.",
    title = "{Short Range Part of Baryon Baryon Interaction in a Quark Model. 1. Formulation}",
    doi = "10.1143/PTP.66.556",
    journal = "Prog. Theor. Phys.",
    volume = "66",
    pages = "556--571",
    year = "1981"
}

@article{Oka:1981rj,
    author = "Oka, M. and Yazaki, K.",
    title = "{Short Range Part of Baryon Baryon Interaction in a Quark Model. 2. Numerical Results for S-Wave}",
    doi = "10.1143/PTP.66.572",
    journal = "Prog. Theor. Phys.",
    volume = "66",
    pages = "572--587",
    year = "1981"
}

@article{Guo:2017jvc,
    author = "Guo, Feng-Kun and Hanhart, Christoph and Mei{\ss}ner, Ulf-G. and Wang, Qian and Zhao, Qiang and Zou, Bing-Song",
    title = "{Hadronic molecules}",
    eprint = "1705.00141",
    archivePrefix = "arXiv",
    primaryClass = "hep-ph",
    doi = "10.1103/RevModPhys.90.015004",
    journal = "Rev. Mod. Phys.",
    volume = "90",
    number = "1",
    pages = "015004",
    year = "2018",
    note = "[Erratum: Rev.Mod.Phys. 94, 029901 (2022)]"
}

@article{Liu:2008xz,
    author = "Liu, Xiang and Liu, Yan-Rui and Deng, Wei-Zhen and Zhu, Shi-Lin",
    title = "{Z+(4430) as a D(1)-prime D* (D(1) D*) molecular state}",
    eprint = "0803.1295",
    archivePrefix = "arXiv",
    primaryClass = "hep-ph",
    doi = "10.1103/PhysRevD.77.094015",
    journal = "Phys. Rev. D",
    volume = "77",
    pages = "094015",
    year = "2008"
}

@article{Bardeen:2003kt,
    author = "Bardeen, William A. and Eichten, Estia J. and Hill, Christopher T.",
    title = "{Chiral Multiplets of Heavy - Light Mesons}",
    eprint = "hep-ph/0305049",
    archivePrefix = "arXiv",
    reportNumber = "FERMILAB-PUB-03-071-T",
    doi = "10.1103/PhysRevD.68.054024",
    journal = "Phys. Rev. D",
    volume = "68",
    pages = "054024",
    year = "2003"
}

@article{Yamaguchi:2022oqz,
    author = "Yamaguchi, Yasuhiro and Yasui, Shigehiro and Hosaka, Atsushi",
    title = "{Open charm and bottom meson-nucleon potentials {\`a}~la the nuclear force}",
    eprint = "2206.01921",
    archivePrefix = "arXiv",
    primaryClass = "hep-ph",
    doi = "10.1103/PhysRevD.106.094001",
    journal = "Phys. Rev. D",
    volume = "106",
    number = "9",
    pages = "094001",
    year = "2022"
}

@article{Yang:2019rgw,
    author = "Yang, Bin and Meng, Lu and Zhu, Shi-Lin",
    title = "{Possible molecular states composed of doubly charmed baryons with coupled-channel effect}",
    eprint = "1906.04956",
    archivePrefix = "arXiv",
    primaryClass = "hep-ph",
    doi = "10.1140/epja/s10050-020-00028-9",
    journal = "Eur. Phys. J. A",
    volume = "56",
    number = "2",
    pages = "67",
    year = "2020"
}

@article{Lu:2026iwm,
    author = "Lu, An-Su and Yan, Mao-Jun and An, Chun-Sheng and Deng, Cheng-Rong",
    title = "{Heavy dibaryons {\ensuremath{\Xi}}cc(*){\ensuremath{\Xi}}cc(*) and {\ensuremath{\Xi}}bb(*){\ensuremath{\Xi}}bb(*)}",
    eprint = "2603.09830",
    archivePrefix = "arXiv",
    primaryClass = "hep-ph",
    doi = "10.1103/j351-hnhy",
    journal = "Phys. Rev. D",
    volume = "113",
    number = "7",
    pages = "074003",
    year = "2026"
}
\end{document}